\PassOptionsToPackage{dvipsnames}{xcolor}
\documentclass[sigconf, 10pt]{acmart}
\usepackage{xspace}
\usepackage{amsmath}
\usepackage{amssymb}
\usepackage{mathtools}
\usepackage{bbm}
\usepackage{bm}
\usepackage{float}
\usepackage{enumitem}
\usepackage{hyperref}
\usepackage{graphicx}
\usepackage[caption=false,lofdepth,lotdepth]{subfig}
\usepackage{listings}
\usepackage{courier}
\usepackage[T1]{fontenc}
\usepackage{multirow}
\usepackage{soul}
\usepackage{xcolor}
\usepackage{bbm, bm}
\usepackage{url}
\usepackage{verbatim}
\usepackage{algorithm}
\usepackage{algpseudocode}
\usepackage{array}
\usepackage{tikz}
\usepackage{ctable}
\usepackage{blkarray}
\usetikzlibrary{bayesnet}
\tikzstyle{mybox} = [draw=black, thick, rectangle]

{}

\newlist{myitemize}{itemize}{1}
\setlist[myitemize,1]{label=\textbullet,itemsep=0pt,leftmargin=10pt,parsep=2pt}

\newcommand{\cut}[1]{}
\newcommand{\eat}[1]{}
\newcommand{\commentresolved}[1]{}

\newcommand{\set}[1]{\{#1\}}                    
\newcommand{\setof}[2]{\{{#1}:{#2}\}}
\usepackage{aliascnt}  		

\newaliascnt{lemma}{theorem}				
\aliascntresetthe{lemma}  					
\newaliascnt{conjecture}{theorem}			
\aliascntresetthe{conjecture}  				
\newaliascnt{remark}{theorem}				
              
\aliascntresetthe{remark}  					
\newaliascnt{corollary}{theorem}			
\aliascntresetthe{corollary}  				
\newaliascnt{definition}{theorem}			
\aliascntresetthe{definition}  				
\newaliascnt{proposition}{theorem}			
\aliascntresetthe{proposition}  				
\newaliascnt{example}{theorem}			
\newtheorem{example}[example]{Example}  	
\aliascntresetthe{example}  				

\usepackage{relsize}

\newcommand\ignore[1]{\relax}

\makeatletter
\def\BState{\State\hskip-\ALG@thistlm}
\newcommand{\vast}{\bBigg@{4}}
\newcommand{\Vast}{\bBigg@{5}}
\makeatother

\newcolumntype{M}[1]{>{\centering\arraybackslash}m{#1}}

\newcommand*{\mathcolor}{}
\def\mathcolor#1#{\mathcoloraux{#1}}
\newcommand*{\mathcoloraux}[3]{%
  \protect\leavevmode
  \begingroup
    \color#1{#2}#3%
  \endgroup
}

\chardef\_=`_

\newcommand{\name}{\textsc{Themis}\xspace}
\newcommand{\STAB}[1]{\begin{tabular}{@{}c@{}}#1\end{tabular}}

\newcommand{\matr}[1]{\bm{#1}}
\newcommand{\ie}{\textrm{i.e.}\xspace}
\newcommand{\eg}{\textrm{e.g.}\xspace}

\newcommand{\model}{\mathcal{M}}

\definecolor{mygreen}{rgb}{0,0.6,0}
\definecolor{mygray}{rgb}{0.5,0.5,0.5}
\definecolor{mymauve}{rgb}{0.58,0,0.82}

\lstdefinestyle{myJava}{ %
  language=java,
  mathescape=true,
  backgroundcolor=\color{white},   
  basicstyle=\scriptsize,        
  breaklines=true,                 
  captionpos=b,                    
  commentstyle=\color{mygreen},    
  escapeinside={*@}{@*},         
  keywordstyle=\color{blue},       
  stringstyle=\color{mymauve},     
}
\lstdefinestyle{mySQL}{ %
    mathescape=true,
    language=SQL,
    basicstyle=\small\ttfamily,
    deletekeywords={MIN},
    otherkeywords={LIMIT, GROUP, BY, ORDER, DESC},
    showstringspaces=false
}

\algdef{SE}[DOWHILE]{Do}{doWhile}{\algorithmicdo}[1]{\algorithmicwhile\ #1}%

\setstcolor{red}

\renewcommand\footnotetextcopyrightpermission[1]{}

\begin{document}

\title{Sample Debiasing in the Themis Open World Database System (Extended Version)}
\author{Laurel Orr, Magdalena Balazinska, Dan Suciu}
\email{ljorr1, magda, suciu @cs.washington.edu}
\affiliation{
  \institution{University of Washington}
  \city{Seattle}
  \state{Washington}
}

\begin{sloppypar}
\begin{abstract}
  Open world database management systems assume tuples {\em not in} the database still exist and are becoming an increasingly important area of research. We present \name~\footnote{This is the extended version of Sample Debiasing in the Themis Open World Database System, {\em SIGMOD 2020}.}, the first open world database that automatically rebalances arbitrarily biased samples to approximately answer queries as if they were issued over the entire population. We leverage apriori population aggregate information to develop and combine two different approaches for automatic debiasing: sample reweighting and Bayesian network probabilistic modeling. We build a prototype of \name and demonstrate that \name achieves higher query accuracy than the default AQP approach, an alternative sample reweighting technique, and a variety of Bayesian network models while maintaining interactive query response times. We also show that \name is robust to differences in the support between the sample and population, a key use case when using social media samples.
\end{abstract}

\maketitle

\section{Introduction}
\label{sec:introduction}
Data samples are increasingly easy to access and analyze with the help of websites such as Facebook and Twitter and data repositories such as data.gov, data.world, and kaggle.com. Additionally, data analytic toolkits, like Python, are becoming more mainstream. These two factors have led to data science becoming tightly coupled with sample analysis.

Modern data scientists, however, face the added challenge that the data samples they seek to analyze are not always an accurate representation of the population they are sampled from. For example, social scientists today study migration patterns from Twitter samples~\cite{zagheni2014inferring}, but Twitter users are a non-uniform subset of all people. This phenomenon is known as sample selection bias~\cite{cortes2008sample} and is problematic because it can lead to inaccurate analyses.

Correcting this bias, however, is difficult because the sampling mechanism in today's data sources, \ie the probability of some population tuple being included in the sample, is typically not known. This means common techniques like the Horvitz-Thompson estimator~\cite{beaumont2008new} are not applicable.

There is, however, another increasingly available data source scientists leverage for debiasing: population aggregates. Along with the increase in the number of publicly available data samples, there is a recent push for more data transparency and reporting by corporations and governments, \eg the United State's OPEN Government Data Act passed in 2018~\cite{opendataact} and the InFuse UK aggregate population statistics tool~\cite{infuseUK}. These reports are often in the form of population aggregate queries. For example, in the FBI's 2017 Internet Crime Report~\cite{cybercrimedata}, they present a table showing a \texttt{GROUP BY, COUNT(*)} aggregate query over crime type, counting the number of victims in each crime type group.

These aggregates can facilitate data debiasing, but the process remains tedious and error prone. There is no general, automatic technique or system for debiasing using aggregates. With the ultimate goal of answering queries approximately over the population, data scientists are forced to manually implement one-off, specialized solutions~\cite{zagheni2014inferring} tailored towards specific datasets, such as census reports~\cite{muller2017generalized}.

In this paper, we present \name, which is, to our knowledge, the first open world database management system (OW-DMBS) that automates and encapsulates the debiasing process. The data scientist simply inserts a sample and aggregates and then asks queries, getting approximate results as if the queries were issued on the population. This novel query processing paradigm, which we call open world query processing (OWQP), is fundamentally different from other paradigms where queries are processed over populations (standard query processing) or {\em representative} samples (approximate query processing (AQP) or cardinality estimation).

Supporting OWQP and data debiasing remains unaddressed because DBMSs are traditionally built for closed world data. They only support samples for AQP where DMBSs leverage samples to achieve interactive speeds. However, AQP techniques~\cite{galakatos2017revisiting,ding2016sample+,orr2017probabilistic,mozafari2015handbook,chaudhuri2017approximate,park2017database,peng2018aqp++} make one (or both) of the following assumptions. (1) They have access to the entire population, or (2) they have knowledge of the error from querying the sample. Neither of these assumptions hold in OWQP, making standard AQP techniques not applicable. The default AQP solution is therefore uniform reweighting, which is inaccurate.

As DBMSs become increasingly used by data scientists~\cite{datamangementfuture}, however, they need to support OWQP to meet to the needs of these new users. \name takes the first step in that direction.

To achieve our goal, at the heart of our system, we develop and combine two different debiasing techniques: reweighting the sample and learning the probability distribution of the population. The former allows us to more accurately answer heavy hitter queries while the later ensures we can answer queries about tuples that may not exist in the sample.

For sample reweighting, we investigate two different approaches: modifying linear regression and applying an existing aggregate fitting procedure. For learning the probability distribution, we utilize Bayesian networks to build an approximate population probability distribution. The uniqueness of our system is in not only building two separate debiasing techniques but also combining them into one unified system for query answering.

We build a prototype database system called \name, named after the Greek titan for balance and order. \name treats relations as samples and automatically corrects for sample bias using population-level aggregates. We evaluate \name on three datasets to show that \name is more accurate at answering point queries than the default AQP technique, linear regression reweighting (\autoref{fig:linmodels}), and a variety of Bayesian network probabilistic approaches (\autoref{fig:flights_BN_Hitters}). We further demonstrate that \name can handle more advanced aggregate queries, depending on the apriori knowledge.
In summary, the contributions of this paper are as follows:
\begin{myitemize}
\vspace{-0.3pt}
\item A new query processing paradigm (OWQP).
\item The first OW-DMBS that automatically debiases data from a sample and population aggregates for OWQP (\autoref{sec:model}).
\item The development and application of debiasing techniques and a novel hybrid approach integrating them (\autoref{sec:data_debiasing}).
\item Two optimization techniques for faster preprocessing: aggregate pruning and model simplification (\autoref{sec:optimizaiton}).
\item Detailed experiments on three datasets showing that \name achieves a 70 percent improvement in the median error when compared to the default AQP approach when asking about heavy hitter tuples (\autoref{tab:perc_improvement_HH}, \autoref{sec:evaluation}). We further show \name is robust to differences in the support of the sample and the population.
\end{myitemize}

The paper is organized as follows. \autoref{sec:motivating_example} gives a motivating example, \autoref{sec:model} gives the model of \name, \autoref{sec:data_debiasing} describes our technique in detail, and \autoref{sec:optimizaiton} discusses optimizations. Finally, \autoref{sec:evaluation} provides experimental results.

\section{Motivating Example}
\label{sec:motivating_example}
A data scientist is trying to estimate the number of flights under 30~min in different states of the United States in a year. She has a sample of all flights in the United States biased towards four major states, but she does not know how badly it is biased. Further, she has access to how many flights in total leave from each state. She decides to analyze this data and focus only on short flights on either the East or West Coast of the country. 

Being a database user, she ingests the data into a SQL database. As this dataset is a sample, she has three choices for how to prepare her data for analysis: do nothing, uniformly rebalance (default AQP), or use state information to reweight flights based on the number of flights leaving each state. For the second option, she knows there are 7 million flights in the United States per year but only 700,000 in her sample. Therefore, she adds a \texttt{weight} attribute to the dataset and gives each tuple a weight of 10, indicating the each tuple in her sample represents 10 tuples in the real world. For the third option, if she knows there are $N$ flights leaving from some state per year but only $n$ leaving that state in her sample, she sets the weight of each flight from that state to be $N/n$.

After preprocessing, she starts issuing point queries of the form
\vspace{-5pt}
\begin{lstlisting}[style=mySQL]
SELECT SUM(weight) AS num_flights
FROM flights WHERE flight_time <= 30 min
AND origin_state = `<state>';
\end{lstlisting}
\vspace{-5pt}

The results of a few queries are shown in \autoref{tab:motivating_ex} where Raw represents option one, AQP represents option two, US State represents option three, and \name represents our system's answer. \name and US State use the single aggregate to produce more accurate answers than Raw and AQP because they are correcting for the fact that some flights leaving the four major states are overrepresented in the sample. More importantly, \name does the re-balancing automatically, which will become time consuming to do manually for more complex aggregates. \name also answers queries about tuples not in the sample, like ME. 

\begin{small}
\begin{table}
\centering
\begin{tabular}{|c|c|c|c|c|c|}
\hline
Query & True & Raw & AQP & US State & \name \\ \hline
CA & 7855 & 2846 & 28460 & 7843 & 7843\\ \hline
FL & 2 & 1 & 10 & 3 & 3\\ \hline
OH & 119 & 1 & 10 & 70 &  70\\ \hline
ME & 2 & 0 & 0 & 0 & 3\\ \hline
\end{tabular}
\caption{Query results of the data scientists using the raw sample, a uniformly scaled sample, a state-scaled sample, and \name.}
\label{tab:motivating_ex}
\end{table}
\end{small}

\section{{\large \name} Model}
\label{sec:model}

\begin{figure}[t]
    \centering
    \includegraphics[width=0.9\linewidth]{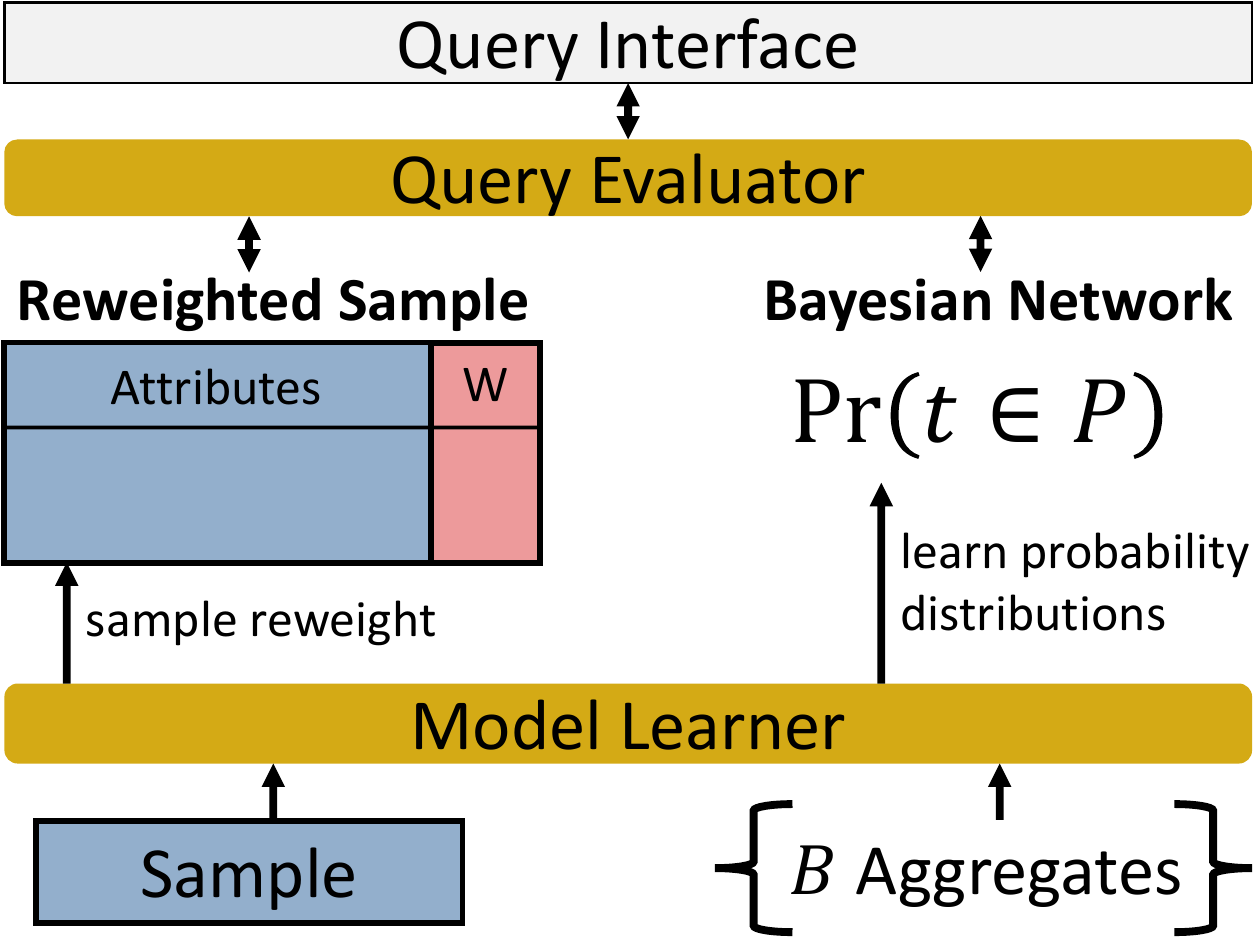}
    \caption{Architecture}
    \label{fig:architecture}
\end{figure}

We now describe our setup and give an overview of \name (see \autoref{fig:architecture}). At a high level, \name uses a sample and population aggregate data to build a model to perform OWQP. We use the term {\em model} because it encapsulates that we use both a reweighted sample and a probabilistic model to answer queries. Both techniques treat the aggregates as constraints to be satisfied.

Note that the population aggregates do not need to be exact. They may contain errors, be computed at different times, or be purposely perturbed. For example, the 2020 US census will add random noise to their reports to be differentially private~\cite{dwork2019differential}. \name will still treat these aggregates as marginal constraints to be satisfied.

We assume there is a well defined, but unavailable population $P$ of (approximate) size $n$ with $m$ attributes $\mathcal{A} = \set{A_1, \ldots, A_m}$. $P$ is unavailable because either it does not exist (\eg a dataset of all graduate students in the US) or is not released to the public (\eg a hospital's private medical data). The active domain of each attribute $A_i$, of size $N_i$, is assumed to be discrete and ordered\footnote{We support continuous data types by bucketizing their active domains.}.

We assume there is a sample $S$ drawn independently but not uniformly from $P$ of size $n_S$ such that for each tuple $t \in P$, $t$ has probability $\Pr_S(t) \geq 0$ of being included in $S$. The subscript $S$ indicates the sampling probability (also called sampling mechanism or propensity score). This probability, however, is not known apriori.

Lastly, we have $\Gamma$, a set of results of $B$ aggregate $\texttt{COUNT(*)}$ queries of various dimensions computed over the population denoted
\begin{equation*}
\label{eq:gamma}
\Gamma = \setof{G_{\matr{\gamma}_i, \textrm{COUNT(*)}}(P)}{i = 1,B}
\end{equation*}
where $G_{\matr{\gamma}_i, \textrm{COUNT(*)}}(P)$ is an aggregate query of dimension $d_i$; \ie, $\matr{\gamma}_i \subseteq \mathcal{A}$ (see~\autoref{ex:one}). Each aggregate query $\Gamma_i$ returns a set of $M_i$ attribute value-count pairs denoted
\begin{equation*}
\label{eq:gamma_i}
\Gamma_i = \setof{(\matr{a}_{i,k}, c_{i,k})}{k = 1, \ldots, M_i}
\end{equation*}
where $\matr{a}_{i,k}$ is the vector of $d_i$ attribute values associated with group $k$ of aggregate $i$, and $c_{i,k}$ is the group's count.

We let $\bigcup_{i = 1,B}\matr{\gamma}_i \subseteq \mathcal{A}$, meaning the aggregates may not cover all domain attributes. When we wish to use vectors of the counts and group values of aggregate $i$ separately, we will use $\overline{\Gamma}_i^C$ and $\overline{\Gamma}_i^A$, respectively.

\begin{example}
\label{ex:one}
Following the example from~\autoref{sec:motivating_example}, assume the population $P$ and sample $S$ are the following sets of domestic flights in the United States. \texttt{date} is the month and \texttt{o\_st} and \texttt{d\_st} are origin and destination states, respectively.
\begin{small}
\begin{center}
\begin{tabular}{c c}
\begin{tabular}{|c|c|c|}
\multicolumn{3}{c}{$P = $} \\
\hline
\texttt{date} & \texttt{o\_st} & \texttt{d\_st} \\ \hline
01 & FL & FL \\ \hline
01 & FL & FL \\ \hline
02 & FL & NY \\ \hline
01 & NC & FL \\ \hline
02 & NC & NY \\ \hline
02 & NC & NY \\ \hline
02 & NC & NY \\ \hline
01 & NY & FL \\ \hline
01 & NY & NC \\ \hline
02 & NY & NY \\ \hline
\end{tabular} 
&
\begin{tabular}{|c|c|c|}
\multicolumn{3}{c}{$S = $} \\
\hline
\texttt{date} & \texttt{o\_st} & \texttt{d\_st} \\ \hline
01 & FL & FL \\ \hline
01 & FL & FL \\ \hline
02 & NC & NY \\ \hline
01 & NY & NC \\ \hline
\end{tabular} \\
\end{tabular}
\end{center}
\end{small}
Let $\Gamma = \set{\Gamma_1, \Gamma_2}$ with $d_1 = 1$ and $d_2 = 2$ be the following two aggregate queries.
\begin{flalign*}
&\Gamma_1 = G_{\texttt{date}, \texttt{COUNT(*)}}(P) = \set{([\textrm{01}], 5), ([\textrm{02}], 5)} \\
&\Gamma_2 = G_{\texttt{o\_st}, \texttt{d\_st}, \texttt{COUNT(*)}}(P) = \\
& \set{([\textrm{FL}, \textrm{FL}], 2), ([\textrm{FL}, \textrm{NY}], 1), ([\textrm{NC}, \textrm{FL}], 1), \\
& ([\textrm{NC}, \textrm{NY}], 3), ([\textrm{NY}, \textrm{FL}], 1), ([\textrm{NY}, \textrm{NC}], 1), ([\textrm{NY}, \textrm{NY}], 1)}.
\end{flalign*}
In this case, $n = 10$, $B = 2$, and the $\Gamma$s are set as
\begin{align*}
\overline{\Gamma}_1^A &= \hspace{0.6cm} \begin{bmatrix} \textrm{01} \\ \textrm{02}\end{bmatrix}& \hspace{1cm}  &\overline{\Gamma}_1^C = \begin{bmatrix} 5 \\ 5\end{bmatrix}\\
\overline{\Gamma}_2^A &= 
\begin{bmatrix}
\textrm{FL} & \textrm{FL} \\
\textrm{FL} & \textrm{NY} \\
\textrm{NC} & \textrm{FL} \\
\textrm{NC} & \textrm{NC} \\
\textrm{NY} & \textrm{FL} \\
\textrm{NY} & \textrm{NC} \\
\textrm{NY} & \textrm{NY} \\
\end{bmatrix}& &\overline{\Gamma}_2^C = \begin{bmatrix}2 \\ 1 \\ 1 \\ 3 \\ 1 \\ 1 \\ 1 \end{bmatrix}.
\end{align*}
The ordering of the rows of $\overline{\Gamma}_i^C$ and $\overline{\Gamma}_i^A$ does not matter, as long as it is consistent.
\end{example}

We are given a user query $Q$ over the population. As we do not have $P$, we need to use $S$ and $\Gamma$ to perform OWQP and estimate $Q(P)$. While $Q$ can be any SQL query, we focus on point\footnote{We define a d-dimensional point query as \texttt{SELECT COUNT(*) FROM R WHERE A1 = v1 AND ... AND Ad = vd}.} and \texttt{GROUP BY} queries to study the improvement in accuracy. To answer $Q(P)$, we build a model $\model(\Gamma, S)$ such that $Q(\model(\Gamma, S))$ is an approximate answer to $Q(P)$.

\section{Data Debiasing}
\label{sec:data_debiasing}
We now present how \name builds $\model(\Gamma, S)$. \name has two components: a reweighted sample and a probabilistic model. We present each technique and then describe how \name merges them into a unique hybrid approach for OWQP.

\subsection{Sample Reweighting}
\label{sec:debias_techniques_sample_based}
In sample reweighting, each tuple $t \in S$ gets assigned a weight $w(t)$ indicating the number of tuples it represents in $P$. Queries get transformed to run on weighted tuples by, for example, translating \texttt{COUNT(*)} to be \texttt{SUM(weight)}. If the sampling mechanism, $\Pr_S(t)$, is known, we can use the Horvitz-Thompson estimator which reweights each tuple by $1/\Pr_S(t)$~\cite{cortes2008sample,mozafari2015handbook}.

The challenge is that we do not have the sampling mechanism. As $P$ does not exist, the default approach used by standard AQP systems is to perform uniform reweighting by setting $w(t)$ to be $|P|/|S|$. When the sample is biased, this achieves low accuracy (see \autoref{sec:evaluation}). To correct for the bias, we present two solutions for learning $w(t)$ using the sample $S$ and aggregate information $\Gamma$. The first is to adapt linear regression, and the second is to apply Iterative Proportional Fitting (IPF)~\cite{idel2016review,lovelace2015evaluating}.

\subsubsection{Linear Regression Reweighting}
Following propensity score research~\cite{rosenbaum1984reducing,mccaffrey2013tutorial,austin2011introduction}, we make the same assumption that a tuple's weight depends on the attributes of $t$. In particular, we assume $w(t)$ is a linear combination of its attributes; \ie, if $\matr{t}^{0/1}$ represents the one-hot encoded tuple $t$, then $w(t) = \matr{\beta} \cdot \matr{t}^{0/1}$ where $\matr{\beta}$ is a vector of weights. Note that for the rest of this section, we use $m$ to refer to the number of attributes covered by the aggregates and only use those attributes for learning the weight.

To solve for $\matr{\beta}$, $S$ gets represented by a $n_S \times m^{0/1}$ matrix, $\matr{X}_S$, where $m^{0/1} = \sum_{i = 1}^m N_i + 1$ (the plus one is because we use the standard formulation of adding a column of ones to represent the intercept). $\matr{y}$ is $\overline{\Gamma}_1^C \oplus \ldots \oplus \overline{\Gamma}_{B}^C$ where $\oplus$ represents row-wise concatenation (equivalent to vertically stacking the vectors). In this case, $\matr{y}$ is a column vector of size $\sum_{i=1}^{B} M_i$ containing all the aggregate queries' count values; \ie
$$
\matr{y}
 =
 \begin{bmatrix}
    c_{1,1} &
    \cdots &
    c_{1,{M_1}} &
    \cdots &
    c_{B,1} &
    \cdots &
    c_{B, M_{B}}
 \end{bmatrix}^{T}.
$$
Let $\matr{X}$ be the matrix product of $\matr{G}^{0/1}\matr{X}_S$ where $\matr{G}^{0/1}$ is a $0/1$ matrix with $\sum_{i=1}^{B} M_i$ rows and $n_S$ columns (see~\autoref{ex:lin_reg2}). $\matr{G}^{0/1}$ is an incidence matrix where row $r$ and column $c$ is 1 if row $c$ of $\matr{X}_S$ participates in the $r$th group by result; \ie, if the $r$th attribute value from $\overline{\Gamma}_1^A \oplus \ldots \oplus \overline{\Gamma}_B^A$ is in row $c$ of $S$. We then solve
\begin{equation}
\label{eq:modified_lin_reg}
[\matr{G}^{0/1}\matr{X}_S]\matr{\beta} = \matr{y}.
\end{equation}
In the case an entire row of $\matr{G}^{0/1}\matr{X}_S$ is all zeros, which happens with missing values in $S$, we drop that row and its associated value in $\matr{y}$.

Departing from standard solving techniques, we solve~\autoref{eq:modified_lin_reg} using a constrained least squares formulation to constrain $\matr{\beta}$ to be strictly positive. This enforces that each tuple in the sample gets $w(t) \geq 0$ and is represented in the population.

Further, as we want to avoid $w(t) = 0$, we add an additional row of $[n_S, 0, \ldots, 0]$ with $\sum_{i=1}^m N_i$ zeros to the matrix $\matr{G}^{0/1}\matr{X}_S$ and add the associated value of $n_S$ to $\matr{y}$. This encourages the intercept value to be positive, which will force every tuple to get some positive weight (since the $\matr{\beta}$ parameters are already positive). Note, as we just want to influence the intercept value, we cannot achieve this by adding a row of ones to $\matr{G}^{0/1}$ because this will result in an additional row of $n_S$ followed $\sum_{i=1}^m N_i$ non-zero values added to $\matr{G}^{0/1}\matr{X}_S$.

\begin{example}
\label{ex:lin_reg2}
Continuing with~\autoref{ex:one}. The one-hot encoded version of $S$, $\matr{X}_S$, is
$$
 \begin{blockarray}{ccccccccc}
 \texttt{1s} & \texttt{d}_{01} & \texttt{d}_{02} & \texttt{o}_{FL} & \texttt{o}_{NC} & \texttt{o}_{NY} & \texttt{d}_{FL} & \texttt{d}_{NC} & \texttt{d}_{NY} \\
 \begin{block}{[ccccccccc]}
    1 & 1 & 0 & 1 & 0 & 0 & 1 & 0 & 0 \\
    1 & 1 & 0 & 1 & 0 & 0 & 1 & 0 & 0 \\
    1 & 0 & 1 & 0 & 1 & 0 & 0 & 0 & 1 \\
    1 & 1 & 0 & 0 & 0 & 1 & 0 & 1 & 0 \\
 \end{block}
 \end{blockarray}.
$$
Our aggregate matrix is
$$
\matr{G}^{0/1}
 =
 \begin{blockarray}{ccccc}
 \begin{block}{[cccc]c}
    1 & 1 & 0 & 1 & \hspace{10pt}\texttt{date} = \textrm{01} \\
    0 & 0 & 1 & 0 & \hspace{10pt}\texttt{date} = \textrm{02} \\
    1 & 1 & 0 & 0 & \hspace{10pt}\texttt{o\_st} = \textrm{FL}\ \&\ \texttt{d\_st} = \textrm{FL}  \\
    0 & 0 & 0 & 0 & \hspace{10pt}\texttt{o\_st} = \textrm{FL}\ \&\ \texttt{d\_st} = \textrm{NY}  \\
    0 & 0 & 0 & 0 & \hspace{10pt}\texttt{o\_st} = \textrm{NC}\ \&\ \texttt{d\_st} = \textrm{FL}  \\
    0 & 0 & 1 & 0 & \hspace{10pt}\texttt{o\_st} = \textrm{NC}\ \&\ \texttt{d\_st} = \textrm{NY}  \\
    0 & 0 & 0 & 0 & \hspace{10pt}\texttt{o\_st} = \textrm{NY}\ \&\ \texttt{d\_st} = \textrm{FL}  \\
    0 & 0 & 0 & 1 & \hspace{10pt}\texttt{o\_st} = \textrm{NY}\ \&\ \texttt{d\_st} = \textrm{NC}  \\
    0 & 0 & 0 & 0 & \hspace{10pt}\texttt{o\_st} = \textrm{NY}\ \&\ \texttt{d\_st} = \textrm{NY}  \\
 \end{block}
 \end{blockarray}
$$
where the right-most column shows the group attributes corresponding to the row in the matrix. After $\matr{G}^{0/1}\matr{X}_S$ is calculated, we add the row of $[4, 0, \ldots, 0]$ at the bottom.
Finally, our solution vector is
$$
\matr{y}
 =
 \begin{bmatrix}
    5 &
    5 &
    2 &
    1 &
    1 &
    3 &
    1 &
    1 &
    1 &
    4
 \end{bmatrix}^{T}
$$
where the final 4 is from adding the $n_S$ constraint to $\matr{y}$.
\end{example}

With these two changes, we solve for $\matr{\beta}$ and $w(t) = \matr{\beta} \cdot \matr{t}^{0/1}$. The final processing step is to modify $w(t)$ so that $\sum_{t \in S} w(t) = n$ by a multiplicative update to each $w(t)$ of $\frac{n}{\sum_{t \in S} w(t)}.$ We do this sum-normalization to correctly reflect to true size of the population {\em after} learning $w(t)$. Note that uniform reweighting is equivalent to setting $w(t) \equiv 1$ before sum-normalizing.

\subsubsection{Iterative Proportional Fitting}
An alternative approach to finding $w(t)$ is to assume every $w(t)$ is independent and can be solved for directly. Inspired by the technique of population synthesis in demography, we apply a technique called Iterative Proportional Fitting (IPF)~\cite{lovelace2015evaluating,beckman1996creating,muller2017generalized,sun2015bayesian,farooq2013simulation,deming1940least} to solve for $w(t)$. While IPF is not new, our approach of using IPF for arbitrary data debiasing is novel.

To briefly review IPF, IPF is a simple iterative procedure for calibrating sample weights to match given population aggregates and is traditionally used to reweight representative microsamples of some population to aggregate census reports. For each individual aggregate, if that aggregate is not satisfied in the sample, the weights of the participating tuples are rescaled to satisfy the selected aggregate. The procedure continues to iterative over aggregates until all aggregates are satisfied. It converges to a satisfactory scaling if such a scaling exists\footnote{IFP is the same algorithm as in matrix scaling, the RAS algorithm, and biproportional fitting~\cite{sinkhorn1967diagonal,lovelace2015evaluating}.}. If no scaling exists, the algorithm may not converge and can only give an approximate reweighting.

The iterative algorithm begins by building the same incidence matrix, $\matr{G}^{0/1}$, as before, where each row represents a single constraint and each column represents a tuple in $S$. $\matr{y}$ is the vector of all aggregate queries' count values. With IPF, however, we have no $\matr{X}_S$. Instead, we have a $n_S$ sized vector $\matr{w}$ of the weights of each tuple; \ie, $\matr{G}^{0/1}\matr{w} = \matr{y}.$ At each iteration, a value in $\matr{w}$ is updated so that its associated aggregate constraint is satisfied.

The pseudocode for IPF is shown in~\autoref{alg:ipf} where $[j]$ represents getting row $j$ for matrices and element $j$ for vectors. At each iteration, if the dot product of the $j$th row of $\matr{G}^{0/1}$ with $\matr{w}$ does not equal the $j$th element in $\matr{y}$, the weights are scaled so that the constraint is satisfied. Note that only the weights participating in the aggregate, \ie, with nonzero $\matr{G}^{0/1}[j]$ values, are updated. 

\begin{small}
\begin{algorithm}[t]
\caption{\textbf{\textsc{IPF}}~\cite{muller2017generalized}}
\label{alg:ipf}
\begin{algorithmic}
\State $\mathit{iter} \gets 1$
\While{not converged or $\mathit{iter} < \mathit{maxIter}$}
    \For {$j = 1$ to $\sum_{i=1}^B M_i$}
        \If {$\matr{G}^{0/1}[j]\cdot\matr{w} \neq \matr{y}[j]$}
            \State $s \gets \frac{\matr{y}[j]}{\matr{G}^{0/1}[j]\cdot\matr{w}}$
            \For {$i = 1$ to $n_S$}
                \If{$\matr{G}^{0/1}[j][i] = 1$}
                    \State $\matr{w}[i] \gets s*\matr{w}[i]$
                \EndIf
            \EndFor
        \EndIf
    \EndFor
    \State $\mathit{iter} \gets \mathit{iter}+1$
\EndWhile
\end{algorithmic}
\end{algorithm}
\end{small}

\autoref{ex:ipf} gives an example of the IPF algorithm and further demonstrates how missing values in $S$ can prevent IPF from converging. 
\begin{example}
\label{ex:ipf}
Take $S$, $\matr{G}^{0/1}$, and $\matr{y}$ as shown in \autoref{ex:lin_reg2}. $\matr{w}$ is size 4, one weight per row of $S$. We show $S$ with $\matr{w}$ after each iteration as an additional column below. IPF iterates over the aggregates in the same order as the rows of $\matr{G}^{0/1}$.

Following the pseudo code, we start with $j = 1$. $\matr{G}^{0/1}[1] = \begin{bmatrix} 1 & 1 & 0 & 1 \end{bmatrix}$ and $\matr{y}[1] = 5$. These represent the aggregate $\texttt{date} = \textrm{01}$ having a count of 5. As $\matr{G}^{0/1}[1]\cdot\matr{w} = 3$, we update $\matr{w}$ so the first, second, and fourth elements are $5/3$. This is shown in the weight column for $j = 1$ below.

When $j = 2$, $\matr{G}^{0/1}[2] = \begin{bmatrix} 0 & 0 & 1 & 0 \end{bmatrix}$ and $\matr{y}[1] = 5$. As $\matr{G}^{0/1}[2]\cdot\matr{w} = 1$, we set the third element of $\matr{w}$ to be $5/1$, as shown in the column for $j = 2$.

\begin{small}
\begin{center}
\begin{tabular}{|c|c|c|c|c|c|c|c|c|}
\cline{4-9}
\multicolumn{3}{r|}{} & \multicolumn{6}{c|}{Weight Values After Iteration} \\ \cline{4-9}
\multicolumn{3}{r|}{$\mathit{iter} = $} & 0 & 1 & 1 & 1 & \ldots & 1 \\ \cline{4-9}
\multicolumn{3}{r|}{$j =$} & 0 & 1 & 2 & 3 & \ldots & 9 \\ \hline
\textbf{\texttt{date} }& \textbf{\texttt{o\_st}} & \textbf{\texttt{d\_st}} & $\matr{w}$ & $\matr{w}$ & $\matr{w}$ & $\matr{w}$ & \ldots & $\matr{w}$ \\ \hline
01 & FL & FL & 1 & 1.67 & 1.67 & 1    & \dots & 1 \\ \hline
01 & FL & FL & 1 & 1.67 & 1.67 & 1    & \dots & 1 \\ \hline
02 & NC & NY & 1 & 1    & 5    & 5    & \dots & 3 \\ \hline
01 & NY & NC & 1 & 1.67 & 1.67 & 1.67 & \dots & 1 \\ \hline
\end{tabular}
\end{center}
\end{small}

The process continues for all 9 rows of $\matr{G}^{0/1}$ until we get the weights after $j = 9$ and $\mathit{iter} = 1$, shown in the last column. We see that at the end , the weights for the tuples with $\texttt{date} = \textrm{01}$ are back to their original value of one. When we go through the process again for $j = 1$ and $\mathit{iter} = 2$, those weights will be scaled back to $5/3$. In this case, IPF will not converge because the sample is missing tuples that fly to and from \texttt{FL}, but IPF does give us an approximate reweighting.
\end{example}
We show in~\autoref{sec:evaluation} that even when IPF does not converge, the approximate weights still achieve high accuracy for queries asking about tuples in $S$.

Lastly, while both the reweighting techniques solve for $w(t)$, that linear regression has $m^{0/1}$ parameters while IPF has $n_S$ parameters. Typically, $m^{0/1} < n_S$. Further, since both methods have $\sum_{i = 1,B}M_i$ constraints, typically, linear regression is over constrained while IPF is under constrained.

\subsection{Probabilistic Model Learning}
\label{sec:debias_techniques_model_based}
We just presented two different reweighting schemes to debias $S$ using the aggregates $\Gamma$. It is important to understand when sample reweighting will fail. For one, the Horvitz-Thompson estimator, which we are approximating by $w(t)$, assumes the support of the sample is the same as the population, \ie $Pr_S(t) > 0 \ \forall t$. When this does not hold, \eg when the sampling design is flawed, sample reweighting is inaccurate. Secondly, even if the support is the same, sample reweighting will fail when tuples exist in $P$ but not in $S$ because the sample will always say those tuples do not exist. This occurs with rare groups and small sample sizes. While we could impute missing rows to $S$, this risks losing important structural information ($S$ gives us partial information about the manifold $P$ lives on) and slowing down queries.

This section presents our solution to this problem: build a probabilistic model of $P$ using $S$ and $\Gamma$ and answer queries using this model~\cite{deshpande2006mauvedb,orr2017probabilistic}\footnote{In order to reason about the population probability distribution, we use the possible world semantics.}. When building a probabilistic model, the first consideration is what class of distributions to use. For example, if the population is believed to be Gaussian in nature, learning a mixture of Gaussians will likely be optimal. As we have no prior knowledge on the population, our main concern is choosing a distribution that can be learned from aggregate data. Similar to~\cite{sun2015bayesian}, we use a Bayesian network (BN) to model the population distribution as a Bayesian network is parameterized by aggregate queries and can scale to many attributes and large data~\cite{friedman1999learningmassive}. Unlike~\cite{sun2015bayesian}, which builds the BN from the sample only, the novelty of our BN framework is that is merges $S$ and $\Gamma$ into BN learning.

\subsubsection{Why Standard Bayesian Network Algorithms Do Not Apply}
\label{sec:debias_techniques_model_based_bayesiannetworks}
\begin{small}
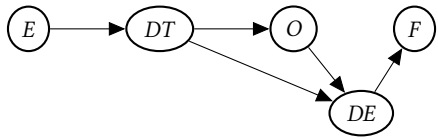
\begin{figure}[t]
  \begin{center}
    \begin{tikzpicture}
      \tikzstyle{nd} = [draw,fill=none,ellipse,thick]
      \node[nd]                                            (E) {\small $\mathit{E}$};
      \node[nd,right=of E]                                 (D) {\small $\mathit{DT}$};
      \node[nd,right=of D]                                 (O) {\small $\mathit{O}$};
      \node[nd,below=of O, xshift=0.9cm, yshift=0.5cm]     (S) {\small $\mathit{DE}$};
      \node[nd,right=of O]                                 (F) {\small $\mathit{F}$};
      \edge {S} {F} ;
      \edge {O} {S} ;
      \edge {D} {O, S} ; %
      \edge {E} {D} ; %
    \end{tikzpicture}
  \end{center}
  \caption{Example Bayesian network of flights in the United States (see \autoref{tab:attr_abbr} for abbreviations).}
  \label{fig:bn_eq}
\end{figure}
\end{small}
Recall that a Bayesian network is a probabilistic graphical model representing a set of random variables and their conditional dependencies through a directed, acyclic graph. Each edge represents a conditional dependency factor of the form $\Pr(X_i|\mathit{Pa}(X_i))$ where $\mathit{Pa}(X_i)$ are the parents of node $X_i$. The example in~\autoref{fig:bn_eq} represents the joint distribution is $\Pr(E,\mathit{DT},O,\mathit{DE},F) = \Pr(\mathit{DT}|E)\Pr(O|\mathit{DT})\Pr(\mathit{DE}|O, \mathit{DT})\Pr(F|\mathit{DE})\Pr(E)$.

Standard BN learning algorithms assume access to the population $P$. They first learn the structure~\cite{campos2006scoring,margaritis2003learning,friedman1999learningmassive} and then the factor parameters by setting them to be the fraction of times each possible state of $X_i$ occurs given the possible states of $\mathit{Pa}(X_i)$.

We fundamentally cannot use standard, black-box BN learning algorithms as we do not have access to $P$. We could learn the BN just using $S$, but that ignores the population aggregate data $\Gamma$. We cannot just use $\Gamma$ as $\Gamma$ may not have information on all the attributes. We need to combine $S$ and $\Gamma$ for the highest accuracy (we evaluate this hypothesis in \autoref{sec:eval_BN}). However, doing so is non-trivial, especially for parameter learning.

To see this, take the flights example from \autoref{fig:bn_eq}. Assume we have some sample $S$ and one 2D population aggregate over \texttt{DE} (destination) and \texttt{DT} (distance). The first step is to learn the structure. Structure learning algorithms are either greedy (altering edges one at a time) or constraint-based (using independence tests to find a satisfying structure). For constraint-based algorithms, it is unclear how to test for independence over the attributes \texttt{DE}, \texttt{DT}, and \texttt{E} as only two of them are covered in the aggregate; therefore, we must use a greedy algorithm. We must modify the greedy algorithm, however, to incorporate information from both the sample and the aggregates, prioritizing the aggregates over the sample when possible. We show our approach in \autoref{sec:debias_techniques_model_based_learning_network_structure}.

Parameter learning raises an even greater challenge than structure learning. Suppose we learn the structure shown in \autoref{fig:bn_eq}. Further, suppose the aggregate gives us that the probability of a flight distance of 500 miles is 0.20. Take the edge from \texttt{E} to \texttt{DT}. How do we learn the parameters for this edge's factor $\Pr(\mathit{DT}|E)$? We cannot learn the parameters from $S$ as $S$ may not have any flights traveling 500 miles. Further, how can we ensure that $\Pr(\texttt{DT} = \textrm{500mi}) = \sum_{\texttt{E}}\Pr(E)\Pr(\texttt{DT} = \textrm{500mi}|E)$ is equal to 0.20? This problem becomes more complex as we add aggregates. We solve this problem in \autoref{sec:debias_techniques_model_based_learning_network_parameters}.

\subsubsection{Learning Network Structure}
\label{sec:debias_techniques_model_based_learning_network_structure}
As mentioned previously, we adapt the greedy hill-climbing algorithm~\cite{getoor2001selectivity,margaritis2003learning}. The traditional hill-climbing algorithm's goal is to find the structure that maximizes some score. At each step of the algorithm, it makes the ``move'' (adding, removing, or reversing a directed edge) that improves the score the most. If the score cannot be further improved, the algorithm terminates.

We modify the algorithm as follows. To focus on learning from the population before the sample, our algorithm runs in two phases: building from $\Gamma$ and building from $S$. As $\Gamma$ represents ground truth information, we want to build as many edges from $\Gamma$ before adding edges from $S$. In the first phase, we make ``moves'' using $\Gamma$ until all attributes from $\Gamma$ are added to the network. Then, if there are any remaining attributes in $S$ not in $\Gamma$, we use $S$ to continue building.

In the $\Gamma$ building phase, we further modify the move selection algorithm to ensure we can score a candidate edge from $X_i$ to $X_j$. As scoring requires computing a group by query over $X_i$, $X_j$, and $\mathit{Pa}(X_i)$, we only consider candidate edges that have the necessary support in $\Gamma$; \ie the attributes $X_i$, $X_j$, and $\mathit{Pa}(X_i)$ appear together in some aggregate. In the $S$ building phase, all edges are allowed.

Our last modification is to ``lock in'' edges that are added from the $\Gamma$ phase, meaning we cannot remove them, because we want to keep all structural knowledge from $\Gamma$ intact. This also prevents overfitting to the sample. 

The pseudocode for our greedy hill-climbing algorithm is shown in~\autoref{alg:hill-climbing}. We use the BIC score because it discourages overly complicated structures that could overfit and does not depend on any prior over the parameters~\cite{margaritis2003learning}. $\mathcal{E}$ is the set of directed edges, $\mathcal{T}$ is the set of conditional probability tables needed to parameterize the network, $\mathcal{P} \in \set{1,2}$ is the phase of the algorithm, and $s$ represents the BIC score. The functions \texttt{CondProbTables} and \texttt{BIC} are the same as in the standard algorithm. The function \texttt{BuildEdges}, shown in~\autoref{alg:build_edges}, determines which moves are allowed; \ie, if an edge has the necessary support.

\begin{small}
\begin{algorithm}[t]
\caption{\textbf{\textsc{GreedyHC}}}
\label{alg:hill-climbing}
\begin{algorithmic}
\State $\mathcal{E} \gets \varnothing$, $\mathcal{N} \gets $ all nodes
\State $\mathit{s} \gets -\infty$, $\mathit{s}' \gets -\infty$, $\mathcal{P} = 1$
\Do
    \If {$\mathcal{P} = 1$} $\mathcal{D} \gets \Gamma$ \Else $\ \mathcal{D} \gets S$
    \EndIf
    \For {$(X_i, X_j) \in \mathcal{N} \times \mathcal{N}$}
        \For {$\mathcal{E}' \in $ \Call{BuildEdges}{$(X_i, X_j), \mathcal{E}, D, \mathcal{P}$}}
            \State $\mathcal{T}' \gets $ \Call{CondProbTables}{$\mathcal{D}$, $\mathcal{E}'$}
            \State $\mathit{t} \gets $ \Call{BIC}{$\mathcal{T}'$, $\mathcal{E}'$}
            \If {$\mathit{t} > \mathit{s}'$}
                $\mathit{s}' \gets \mathit{t}$
           \EndIf
        \EndFor
    \EndFor
    \If {$\mathit{attrs}(\Gamma) \in \mathcal{E}$ and $|\mathit{s} - \mathit{s}'| \leq 0$} $\mathcal{P} \gets 2$ \EndIf
\doWhile{$|\mathit{s} - \mathit{s}'| > 0$}
\State \Return $\mathcal{E}$, $\mathcal{T}$
\end{algorithmic}
\end{algorithm}
\end{small}

\begin{small}
\begin{algorithm}[t]
\caption{\textbf{\textsc{BuildEdges}}}
\label{alg:build_edges}
\begin{algorithmic}
\If {$\mathcal{P} = 2$ and $(X_i, X_j)$ added when $\mathcal{P} = 1$}
  \State $\mathcal{S} \gets \set{\mathcal{E} - (X_i, X_j)}$
\EndIf
\If {$\set{X_j, X_i, \mathit{Pa}(X_j)} \in \mathit{attrs(\mathcal{D})}$}
  \State $\mathcal{S} \gets \mathcal{S} \cup \mathcal{E} \cup (X_i, X_j)$
\EndIf
\If {$\set{X_j, X_i, \mathit{Pa}(X_i)} \in \mathit{attrs(\mathcal{D})}$}
  \State $\mathcal{S} \gets \mathcal{S} \cup \mathcal{E} - (X_i, X_j) \cup (X_j, X_i)$
\EndIf
\State \Return $\mathcal{S}$
\end{algorithmic}
\end{algorithm}
\end{small}

\subsubsection{Learning Network Parameters}
\label{sec:debias_techniques_model_based_learning_network_parameters}
As mentioned previously, it is non-trivial how to learn the parameters using $S$ and $\Gamma$. Recall that BN parameters are learned by maximizing the likelihood of the data subject to the BN structure. Inspired by~\cite{de2008constrained,niculescu2005exploiting} adding parameter sharing constraints to the BN parameter learning optimization, we instead add constraints enforcing each aggregate is satisfied. To our knowledge, we are the first to add aggregate constraints to BN parameter learning.

Specifically, let $\mathcal{J}_i = \set{j_{i,1}, \ldots, j_{i,d_i}} \subseteq \set{1, \ldots, m}$ be the attribute index set of aggregate $i$; \ie $\matr{a}_{i,k} = [a_{k,j_{i,1}}, \ldots, a_{k,j_{i,d_i}}]$. Then, for some $(\matr{a}_{i',k'}, c_{i', k'}) \in \Gamma_{i'}$ (we use $i'$ and $k'$ to differentiate from the Bayesian network $i$ and $k$ variables), we add the constraint

$$\sum_{\matr{v} \in \bigtimes_{j' \in \lnot\mathcal{J}_{i'}} \mathit{dom}(A_{j'})} \Pr(X_{\mathcal{J}_{i'}} = \matr{a}_{i', k'}, X_{\lnot\mathcal{J}_{i'}} = \matr{v}) = \frac{c_{i', k'}}{n}$$

to the optimization where $\mathit{dom}$ is the active domain of an attribute, $\bigtimes$ is the cross product, $\lnot\mathcal{J}_{i'} = \set{1,\ldots,m} - \mathcal{J}_{i'}$, $X_{\mathcal{J}_{i'}} = \matr{a}_{i', k'}$ stands for $X_{j_{i',1}} = a_{k',j_{i',1}}, \cdots, X_{j_{i',d_{i'}}} = a_{k',j_{i',d_{i'}}}$, and similarly for $X_{\lnot\mathcal{J}_{i'}} = \matr{v}$. Intuitively, we are summing over all possible values of the attributes, $X_{\lnot\mathcal{J}_{i'}}$, that do {\em not} participate in the aggregate.

For example, following \autoref{fig:bn_eq}, suppose we know that one aggregate attribute-value pair is that 0.2 percent of flights have $\mathit{O} = \texttt{KA}$, $\mathit{DE} = \texttt{NM}$, and $\mathit{ET} = \texttt{60}$. Letting $\theta_{i,j,k}$ represent $\Pr(X_i = j|\mathit{Pa}(X_i) = k)$, the added constraint from that aggregate is
\begin{align*}
\sum_{\mathit{dt} \in \mathit{dom}(\mathit{DT})}\sum_{\mathit{f} \in \mathit{dom}(\mathit{F})} & \theta_{\mathit{DT},\mathit{dt},\set{\texttt{60}}} * \theta_{\mathit{O},\texttt{KA},\set{\mathit{dt}}} \\
 & * \theta_{\mathit{DE},\texttt{NM},\set{\texttt{KA}, \mathit{dt}}} * \theta_{\mathit{F},\mathit{f},\set{\texttt{NM}}} * \theta_{\mathit{E},\texttt{60},\emptyset} = 0.2
\end{align*}

We now get the constrained optimization problem in~\autoref{eq:constrained_optimization} where the first two are standard BN constraints. $\theta_{i,j,k}$ for a particular $i$ and tuple $t$ means $j = t.A_i$ and $k = \setof{t.A_{i'}}{X_{i'} \in \mathit{Pa}(X_i)}$. $\matr{v}$ is the same as before in the sum over all possible values of the attributes that do not participate in the aggregates. $\matr{v}_{j'}$ stands for the individual value of attribute $A_{j'}$. The $*$ stands for the value of the parent, and it will be set to a value in $\matr{a}_{i',k'}$ or $\matr{v}$, depending on if it participates in the aggregate or not.

\begin{alignat}{2}
\label{eq:constrained_optimization}
  & \text{minimize: } -\sum_{t \in S}\sum_{i=1}^{m} \log \theta_{i,j,k} \\
  &
      \nonumber
      \begin{aligned}[t]
      \text{subject to: } \hspace{70pt} \theta_{i,j,k} & \geq 0 & \forall & i,j,k\\
      \sum_{j} \theta_{i,j,k} & = 1 & \quad \forall & i,k\\
      \sum_{\matr{v}}\prod_{j' \in \lnot\mathcal{J}_{i'}}\theta_{A_{j'},\matr{v}_{j'}, *}\prod_{j \in \mathcal{J}_{i'}}\theta_{A_{j},a_{k',j},*} & = \frac{c_{i', k'}}{n} & \quad \forall & (\matr{a}_{i',k'}, c_{i', k'})
      \end{aligned}
\end{alignat}


\subsubsection{Query Answering}
\label{sec:debias_techniques_model_based_query_answering}
Once we have learned the probability distribution of our population, we can answer selection (point) queries probabilistically by calculating $n*\Pr(X_1 = x_1, \ldots, X_m = x_m)$. To answer \texttt{GROUP BY} queries, we use the BN to generate $K$ samples of data that are representative of the population via forward/logic sampling~\cite{sun2015bayesian,korb2010bayesian,henrion1988propagating}. Once the samples $S'_k$ are generated, tuples are uniformly scaled up (\ie the weight of each tuple is $|P|/|S'_k|$), and the query is answered as it is for reweighted samples. After receiving $K$ answers, we return the groups appearing in all $K$ answers, averaging the aggregate value. Using $K$ samples reduces the variance and the number of incorrect ``phantom groups'' (groups that are returned but do not exist).

\subsection{Hybrid Query Evaluator}
\label{sec:hybrid_approach}
To perform OWQP, \name's hybrid approach integrates the previous two methods into a unified technique. For point queries, when a point query gets issued, if the tuple being queried is in the sample, we use the reweighted sample. Otherwise, we do direct BN inference. For \texttt{GROUP BY} queries, we return all values from our reweighted sample unioned with any groups that appear in the BN query but not the reweighted sample query.

The motivation for these techniques is due to the inherent problems with sample reweighting. If the tuple does not exist in the sample, the sample achieves poor accuracy. We are simply capturing this failure in our query evaluator by only using the BN answer to handle missing tuples or groups. The technique is critical when handling samples do not have the same support as the population, as shown in \autoref{sec:evaluation}.

\section{Optimization}
\label{sec:optimizaiton}
There are two main challenges to implementing our techniques efficiently: the potentially large number of aggregates\footnote{InFuse \cite{infuseUK} has more that 900 aggregate queries.} and the nonlinearity of our BN constraints (\autoref{eq:constrained_optimization}). In regards to the former, each new aggregates adds a new constraint in our BN and linear regression solver and is one more iteration of weight rescaling in IPF. We need to reduce this number for efficient solver times. With the later problem, it is well known that nonlinear constraints add complexity and makes solving computationally expensive~\cite{solow2007linear}. In our framework, there are $O(\prod_{j \in \lnot \mathcal{J}_i}N_j)$ variables for {\em each} $(\matr{a}_{i,k}, c_{i,k})$, and without simplifying \autoref{eq:constrained_optimization}, solving is intractable (see \autoref{sec:evaluation}).

To solve these problems, we present two optimization techniques: pruning the least informative aggregates and simplifying the constrained optimization (\autoref{eq:constrained_optimization}). Simplifying \autoref{eq:constrained_optimization} is critical to the success of integrating $S$ and $\Gamma$ into parameter learning as it make solving tractable. Further, it allows for optimizing each BN factor independently.

\subsection{Aggregate Selection}
\label{sec:optimization_aggregate_selection}
Our goal is to reduce the number of aggregates, \ie $|\Gamma|$, {\em before} using them in reweighting and Bayesian network learning. Given a budget $B$, the natural choice is to choose the $B$ most informative aggregates; \ie, the $B$ aggregates that minimize the distance between the true population distribution and some approximate distribution parametrized by the aggregates. We choose to minimize the Kullback-Leibler (KL) divergence because if we assume, like BNs, that our approximate distribution is a product distribution, we can use the fact that Chow-Liu trees~\cite{chow1968approximating} (a second-order product approximate) and their higher order counterparts, a k-order t-cherry junction tree~\cite{bukszar2001probability,szantai2013discovering}, minimize the KL divergence.

Before defining a {\em k-order t-cherry junction tree}, recall that a {\em junction tree}~\cite{wainwright2008graphical,proulx2014modeling} is a tree structure where a node (cluster) is defined by a subset of random variables, denoted $\matr{X}_C$, and is associated with the distribution $\Pr(\matr{X}_C)$. Every tree edge is called a separator and defined by the subset of random variables, denoted $\matr{X}_S$, contained in the intersection of the two clusters being linked. A junction tree must also satisfy the running intersection property that all tree nodes containing the variable $X$ must form a connected region and have every random variable contained in some cluster.

A {\em k-order t-cherry junction tree} is a junction tree with the added properties that (1) every cluster (tree node) consists of exactly $k$ random variables, and (2) every separator consists of exactly $k-1$ random variables. Assuming access to $P$, the greedy algorithm for building a t-cherry tree is to score all possible cluster-separator pairs by $I(\matr{X}_C) - I(\matr{X}_S)$ where $I(\matr{X}_C) = \sum_{i \in C}H(X_i) - H(\matr{X}_C)$ is information content and $H$ is entropy. It greedily adds new cluster-separator pairs with the highest scores as long as, at each iteration, a new random variable is being covered and the separator to be added is contained in an already added cluster. The algorithm terminates once all random variables are covered.

\begin{algorithm}[t]
\caption{Modified t-cherry tree.}
\label{alg:t-cherry-tree}
\begin{algorithmic}
\State $\mathcal{C} \leftarrow $ sorted(\Call{GenClusterSeparatorPairs}{})
\State $\mathcal{C'} \leftarrow \{$ highest scored cluster-separator in $\mathcal{C}$ $\}$
\While {$|\mathcal{C'}| < B$}
    \For {$c \in \mathcal{C}$}
        \If {$c$'s separator contained in some cluster $\in$ $\mathcal{C'}$ 
        \State \hspace{-0.6cm} \textbf{and} new attribute covered by $c$}
          \State add $c$ to $\mathcal{C'}$ 
        \EndIf
    \EndFor
    \If {all attributes covered}
        \State start new tree
        \State $\mathcal{C} \leftarrow $ sorted(\Call{GenClusterSeparatorPairs}{} - $\mathcal{C'}$)
    \EndIf
\EndWhile
\State \Return $\mathcal{C'}$
\end{algorithmic}
\end{algorithm}

Similar to the problem with black-box BN techniques, we cannot use k-order t-cherry junction tree algorithms as they assume access to the entire population. Therefore, we modify the algorithm is two ways (see \autoref{alg:t-cherry-tree}). First, we only initialize cluster-separator pairs that have support in $\Gamma$, meaning we can calculate the mutual information from $\Gamma$ alone. Second, as our aggregate budget $B$ may be larger than the number of attributes, we allow for multiple iterations of our algorithm. To avoid creating duplicate clusters, we disallow previous cluster-separator pairs from being added. Once our algorithm generates $B$ edges, we filter $\Gamma$ so that each $\matr{\gamma}_i$ must be equal to the attributes associated with one of the clusters.

\subsection{Bayesian Network Simplification}
\label{sec:optimization_constrained_optimization}
Our critical and novel optimization is to simplify our constrained optimization from \autoref{eq:constrained_optimization}. To make solving tractable, we want each BN factor $\Pr(X_i | \mathit{Pa(X_i)})$ to be optimized independently with {\em linear} constraints. To do this, we enforce a topological solving order (every parent node is optimized before its children nodes) and limit the aggregates added to our model.

To enforce linear constraints and independent solving, we restrict our model to only add aggregate constraints that act on single factors, \ie aggregate constraints over a child node $X_i$ and its parents. This means for child node $X_i$, the constraints in \autoref{eq:constrained_optimization} will only contain the the product of the child parameter $\theta_{i,j,k}$ with its ancestors because the other factors have marginalized out. By itself, this has only reduced the number of product factors. The key is topological solving order. By insuring that the parents are solved for before the children, at the time of solving for $\theta_{i,j,k}$ for a particular $X_i$, the ancestor terms are already known and become a constant in the constraint, meaning the $\theta_{i,j,k}$ for $X_i$ are the only parameters. By removing aggregate constraints that act on multiple different BN factors, we can turn our nonlinear constraints into linear ones. Further, as we only include constraints on single factors and only those factor's parameters are unknown, we can solve factors independently.

\begin{example}
\label{ex:constrained_optimizaiton_optimized}
Take the example network from \autoref{fig:bn_eq}. Once some factor $\theta_{i,j,k}$ is solved, we denote it $\overline{\theta}_{i,j,k}$. Suppose we have two aggregates over $E$ and $(O, \mathit{DE})$. A topological ordering of all nodes is $E$, $\mathit{DT}$, $O$, $\mathit{DE}$, $F$. To simplify indexing, instead of using $i'$, $k'$ to index the aggregates, we will use the associated values as indexes for the counts. For example, if one aggregate of $(O, \mathit{DE})$ is $(\matr{a}_{2,k'}, c_{2,k'}) = ([\texttt{WI}, \texttt{MN}], 10)$, we write this as $c_{2,\set{\texttt{WI}, \texttt{MN}}} = 10$ (2 representing the second aggregate).

We solve for $E$ first by solving
\begin{alignat*}{2}
  & \text{minimize: } & & -\sum_{t \in S} \log \theta_{E,j,\emptyset} \\
  & \text{subject to: }& \quad & \nonumber
      \begin{aligned}[t]
      \theta_{E,j,\emptyset} & \geq 0 & \forall & j\\
      \sum_{j} \theta_{E,j,\emptyset} & = 1 & \quad &\\
      \theta_{E,j,\emptyset} & = c_{1, j}/n & \quad \forall & j \in \mathit{dom}(\mathit{E}).
      \end{aligned}
\end{alignat*}
Note that because of marginalization of the BN factors, the aggregate constraints do not sum over all possible values of $(\mathit{DT}, O, \mathit{DE}, F)$. Once this optimization is solved, we have $\overline{\theta}_{E,j,\emptyset}$. We solve our next node, $\mathit{DT}$, in closed form because we have no constraints over $\mathit{DT}$. We solve $\mathit{O}$ next by

\begin{alignat*}{2}
  & \text{minimize: }  -\sum_{t \in S} \log \theta_{\mathit{O},j,k} \\
  & \nonumber
      \begin{aligned}[t]
      \text{subject to: } \hspace{30pt} \theta_{\mathit{O},j,k} & \geq 0 & \forall & j,k\\
      \sum_{j} \theta_{\mathit{O},j,k} & = 1 & \forall & k\\
      \sum_{k \in \mathit{dom}(\mathit{DT})} \theta_{\mathit{O},j,k} \sum_{\ell \in \mathit{dom}(\mathit{E})} & \overline{\theta}_{E,\ell,\emptyset}\overline{\theta}_{DT,k,\ell} &&\\
      & = \sum_{v \in \mathit{dom}(\mathit{DE})}\frac{c_{2, \set{j,v}}}{n} & \forall & j \in \mathit{dom}(\mathit{O}).
      \end{aligned}
\end{alignat*}
Note we turn the constraint over $(O, \mathit{DE})$ to be one just over $O$ by aggregation. We use $(O, \mathit{DE})$ again when solving for $DE$. $F$ is solved in closed form.
\end{example}
We can further improve efficiency by limiting the number of parents nodes each child can have in Bayesian network structure learning by modifying the hill-climbing algorithm to prevent adding/reversing edges if a child already has enough parents.

\section{Evaluation}
\label{sec:evaluation}
In this section, we evaluate the accuracy and execution time of \name for OWQP. We compare \name's hybrid approach to the standard AQP solution (uniform reweighting), sample reweighting, and Bayesian network generation. We briefly discuss \name in comparison an AQP approach~\cite{galakatos2017revisiting} that can be modified to leverage aggregates, but as their technique makes different assumptions and we were unable to get code from the authors, we only evaluate one of their techniques. We then investigate the performance of the two different sample reweighting techniques and of the Bayesian network learning technique. Lastly, we demonstrate the benefits of using our pruning technique. We do not show timing results for our optimization from \autoref{sec:optimization_constrained_optimization} as experiments did not finish in under 10 hours without using the optimization, indicating the necessity of the optimizations in making learning tractable.

\subsection{Implementation}
We implemented \name in three parts: the sample reweighter, the BN learner, and the query evaluator (code at~\cite{sourcecode}). The linear regression, IPF, and BN constraint solving were implemented in Python 3.7\footnote{Since the constraint solving was approximate, occasionally a network parameter, \ie $\theta_{i,j,k}$, would be set to some very small negative number. We set these parameters to zero.}. The BN structure learning and inference were implemented in R 3.2 using the BNLearn and gRain package (gRain for exact inference). Lastly, we limited our Bayesian networks to be trees to limit the number of tuning parameters to evaluate. For BN query answering, we used $K=10$.

After learning, the samples with weights stored as an additional column were stored and queried in a Postgres 9.5 database. We performed all experiments on a 64bit Linux machine running Ubuntu 16.04.5. The machine has 120 CPUs and 1 TB of memory. The Postgres database also resides on this machine with a shared buffer size of 250 GB.

\subsection{Datasets}
\label{sec:eval_datasets}
We use a flights dataset~\cite{flightsdata} (all United States flights in 2005 with $n = 6,992,839$), an IMDB dataset~\cite{leis2015good} (actor-movie pairs released in the United States, Great Britain, and Canada with $n = 846,380$), and a synthetic CHILD Bayesian network dataset generated using BNLearn~\cite{bnlearn_repo} with $n = 20,000$. We preprocess the datasets to remove null values and bucketize the real-valued attributes into equi-width buckets. For the two real-world datasets, the attributes and attribute abbreviates are shown in~\autoref{tab:attr_abbr}.

\begin{table}
\begin{footnotesize}
    \centering
    \begin{tabular}{|c|c|}
    \hline
    \textbf{\texttt{Flights}} & \textbf{Abrv} \\ \hline
    \texttt{fl\_date} & \texttt{F} \\ \hline
    \texttt{origin\_state} & \texttt{O} \\ \hline
    \texttt{dest\_state} & \texttt{DE} \\ \hline
    \texttt{elapsed\_time} & \texttt{E} \\ \hline
    \texttt{distance} & \texttt{DT} \\ \hline
    \end{tabular}
    \begin{tabular}{|c|c|}
    \hline
    \textbf{\texttt{IMDB}} & \textbf{Abrv} \\ \hline
    \texttt{movie\_year} & \texttt{MY} \\ \hline
    \texttt{movie\_country} & \texttt{MC} \\ \hline
    \texttt{name} & \texttt{N} \\ \hline
    \texttt{gender} & \texttt{G} \\ \hline
    \texttt{actor\_birth} & \texttt{B} \\ \hline
    \texttt{rating} & \texttt{RG} \\ \hline
    \texttt{top\_250\_rank} & \texttt{TR} \\ \hline
    \texttt{runtime} & \texttt{RT} \\ \hline
    \end{tabular}
\end{footnotesize}
\caption{\texttt{Flights} and \texttt{IMDB} attributes.}
\label{tab:attr_abbr}
\end{table}

We take three samples from \texttt{Flights}: uniform (Unif), flight month of June (June), and flights leaving from a four corner state of CA, NY, FL, WA (SCorners)\footnote{The S stands for the supported Corners sample.}. Each are 10 percent samples with a 90 percent bias, meaning 90 percent of the rows are from the selection criteria. We also take a corner states 10 percent sample with 100 percent bias (Corners).

It is important to understand the motivation for Corners compared to SCorners. Corners represents generating a sample by performing a selection on the population, a common use case. For example, US social media data is a 100 percent biased sample of the population of the US. Only users who have a social media account are in these samples. Because it is common for datasets on the web to be 100-percent biased (\eg \cite{nycflightsR,indianatraffic,baltimoredata}) yet still serve as foundations for analysis, \name needs to handle queries on these types of samples.

We likewise take three samples from \texttt{IMDB}: uniform (Unif), movie country of Great Britain (GB), and movies with ratings 1, 5, or 9 (SR159). We similarly give these 10 percent samples a 90 percent bias and take a 10 percent sample with 100 percent bias of the ratings sample (R159).

As the CHILD data is used to examine our pruning technique, we just use a 10 percent uniform sample.

\setlength{\abovecaptionskip}{5pt plus 3pt minus 3pt}
\setlength{\textfloatsep}{10pt plus 2pt minus 4pt}
\setlength{\floatsep}{10pt plus 2pt minus 2pt}
\subsection{Experimental Setup}
\label{sec:exp_setup}
As real population reports typically have aggregates of one, two, or three dimensions (\eg Excel tables), we use $d = 1$, $2$, or $3$. Note that as the dimensionality of all aggregates is the same, we use $d$ rather than $d_i$. We prune all possible aggregates by our pruning technique to produce from $B = 1$ to $4$ aggregates. \autoref{tab:data_agg} shows the aggregates chosen. For \texttt{IMDB}, we only consider aggregates from the attributes \texttt{MY}, \texttt{MC}, \texttt{G}, \texttt{RG}, \texttt{RT} to investigate the impact of aggregates that do no cover all attributes.
\begin{table}
\begin{small}
    \centering
    \begin{tabular}{|c|c|c|c|}
    \hline
    $d$ & $B$ & \texttt{Flights} & \texttt{IMDB} \\ \hline
    \multirow{4}{*}{2} & 1 & \texttt{E} \& \texttt{DT} & \texttt{MY} \& \texttt{RT} \\ \cline{2-4}
                       & 2 & \texttt{DE} \& \texttt{DT} & \texttt{RG} \& \texttt{RT} \\ \cline{2-4}
                       & 3 & \texttt{O} \& \texttt{DT} & \texttt{MY} \& \texttt{MC} \\ \cline{2-4}
                       & 4 & \texttt{F} \& \texttt{DE} & \texttt{MY} \& \texttt{G} \\ \hline
    \multirow{4}{*}{3} & 1 & \texttt{O} \& \texttt{DE} \& \texttt{DT} & \texttt{MY} \& \texttt{RG} \& \texttt{RT} \\ \cline{2-4}
                       & 2 & \texttt{O} \& \texttt{E} \& \texttt{DT} & \texttt{MY} \& \texttt{MC} \& \texttt{RT} \\ \cline{2-4}
                       & 3 & \texttt{F} \& \texttt{E} \& \texttt{DT} & \texttt{MY} \& \texttt{G} \& \texttt{RT} \\ \cline{2-4}
                       & 4 & \texttt{DE} \& \texttt{E} \& \texttt{DT} & \texttt{MC} \& \texttt{RG} \& \texttt{RT} \\ \hline
    \end{tabular}
    \caption{The 4 2D and 3D \texttt{Flights} and 4 2D and 3D \texttt{IMDB} data aggregate attributes chosen by the pruning technique.}
    \label{tab:data_agg}
\end{small}
\end{table}

To measure accuracy, we run point queries\footnote{We define a d-dimensional point query as \texttt{SELECT COUNT(*) FROM R WHERE A1 = v1 AND ... AND Ad = vd}.} where the query selection values are selected from the population's light hitters (smallest values), heavy hitters (largest values), and random values (any existing value). We run 100 point queries for each of the three selections per attribute set. For \texttt{Flights}, we issue point queries over all possible attribute sets of size two to five (total of 26). For \texttt{IMDB}, as there are too many attributes to run all possible point queries, we randomly choose 20 three dimensional attribute sets\footnote{We use all attributes, not just those covered by aggregates.}. Lastly, we use the error metric of percent difference, $2*|\mathit{true\mathunderscore value} - \mathit{est\mathunderscore value}|/|\mathit{true\mathunderscore value} + \mathit{est\mathunderscore value}|$ rather than percent error to avoid over emphasizing errors where the true value is small and to ensure missed (not in the result but should exist) and phantom (in the result but should not exist) groups get the maximum error of 200 percent.

We also investigate how \name performs for more advanced SQL aggregate queries. We run the six SQL queries shown in \autoref{tab:sql_queries} and measure the average percent difference across the returned groups.

Finally, when measuring runtime (\autoref{sec:eval_runtime}), as all reweighted samples are stored and accessed the same, we only look at the runtime for one reweighted sample.

\subsection{Overall Accuracy}
\label{sec:eval_overall_accuracy}
Using $B = 4$ and $d = 2$, we compare \name's hybrid approach (pink) to the standard AQP approach (uniform reweighting), best linear reweighting technique of IPF (orange), and the best Bayesian network technique of BB (blue) (BB means it uses both $\Gamma$ and $S$ to learn the BN).

\begin{figure}[t]
    \centering
    \includegraphics[width=0.9\linewidth]{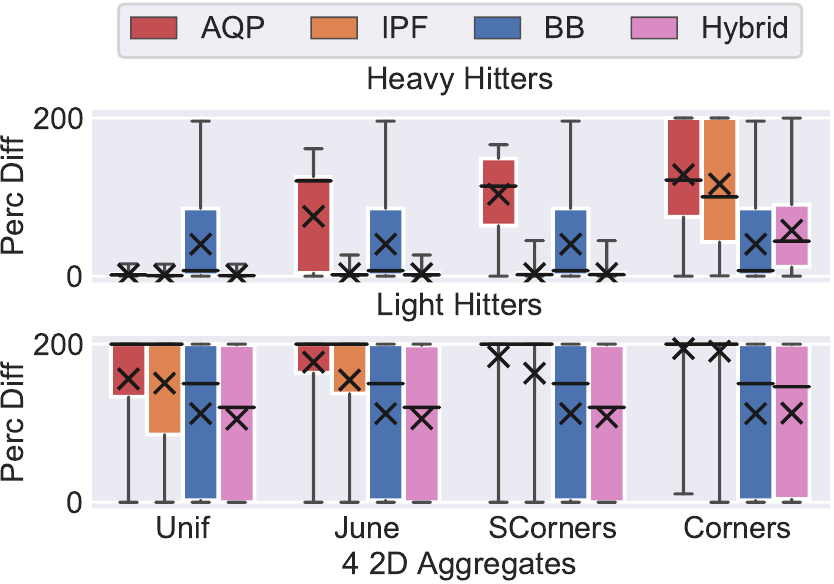}
    \caption{100 heavy and light hitter point query percent difference for \texttt{Flights} biased samples ($B = 4$).}
    \label{fig:flights_sample_Hitters}
\end{figure}

\begin{figure}[t]
    \centering
    \includegraphics[width=0.9\linewidth]{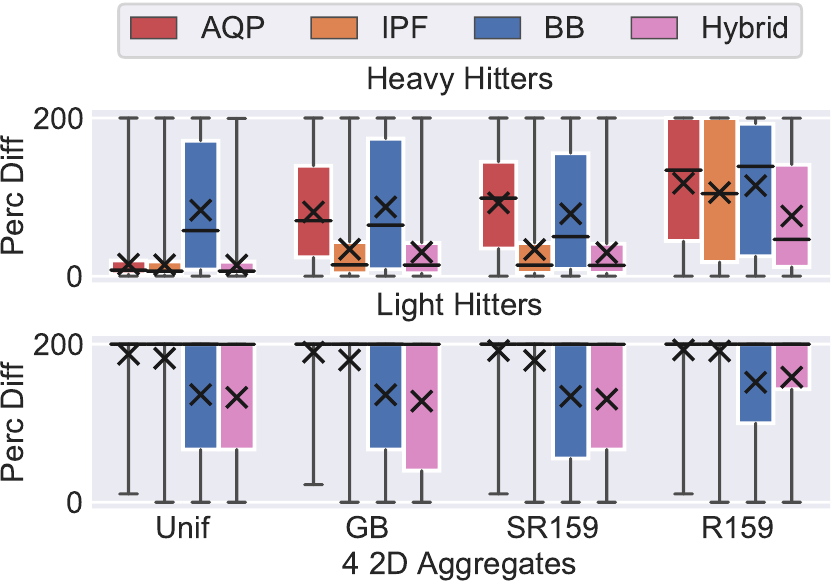}
    \caption{100 heavy and light hitter point query percent difference for \texttt{IMDB} biased samples ($B = 4$).}
    \label{fig:imdb_sample_Hitters}
\end{figure}

\autoref{fig:flights_sample_Hitters} and \autoref{fig:imdb_sample_Hitters} show boxplots of the percent difference of 100 heavy and 100 light hitter point queries across the samples. The median value is the black line and the average is the black X. For reference, \autoref{tab:perc_improvement_HH} shows the percent improvement of the 25th, 50th, and 75th percentiles of \name's hybrid approach to uniform reweighting for \texttt{Flights}.

\begin{small}
\begin{table}
\centering
  \begin{tabular}{|c|l|r|r|r|r|}
    \hline
     \textbf{Hitters} & \textbf{Percentile} & Unif & June & SCorners & Corners \\ \hline
     \multirow{3}{*}{\STAB{\rotatebox[origin=c]{0}{Heavy}}}
     & 25 & 4.2 & 13.6 & 168.3 & 6.1 \\
     & 50 & 1.8 & 69.7 & 61.9 & 2.7 \\
     & 75 & 1.4 & 29.6 & 34.4 & 2.2 \\ \hline
     \multirow{3}{*}{\STAB{\rotatebox[origin=c]{0}{Light}}}
     & 25 & $\infty$ & $\infty$ & $\infty$ & 45 \\
     & 50 & 1.7 & 1.7 & 1.7 & 1.4 \\
     & 75 & 1.0 & 1.0 & 1.0 & 1.0 \\ \hline
  \end{tabular}
\caption{Percent improvement of percentiles for hybrid compared to AQP for the queries from~\autoref{fig:flights_sample_Hitters}. The infinite value represents that hybrid has zero error.}
\label{tab:perc_improvement_HH}
\end{table}
\end{small}

We see that for the samples that have the same support as the population (first three), \name's hybrid technique achieves the lowest error for heavy and light hitters. For the \texttt{Flights} sample without support (Corners), the BN technique (BB) performs best, but hybrid performs better than IPF, indicating that hybrid mitigates the problem of mismatching support, a key requirement for \name's applicability for real world use cases. For light hitters, BB performs better than IPF and AQP, and it is because of this that \name's hybrid approach achieves the lowest error for light hitters. \name uses IPF in the rare case that the tuple is in the sample, which is why hybrid achieves lower error than BB.

BB does not perform best for R159 because of queries over the very dense attribute \texttt{N} (48,000 distinct values). BB learns that \texttt{N} is uniformly distributed and underestimates queries over \texttt{N} because all values are equally likely. 

\begin{figure}[t]
    \centering
    \includegraphics[width=0.9\linewidth]{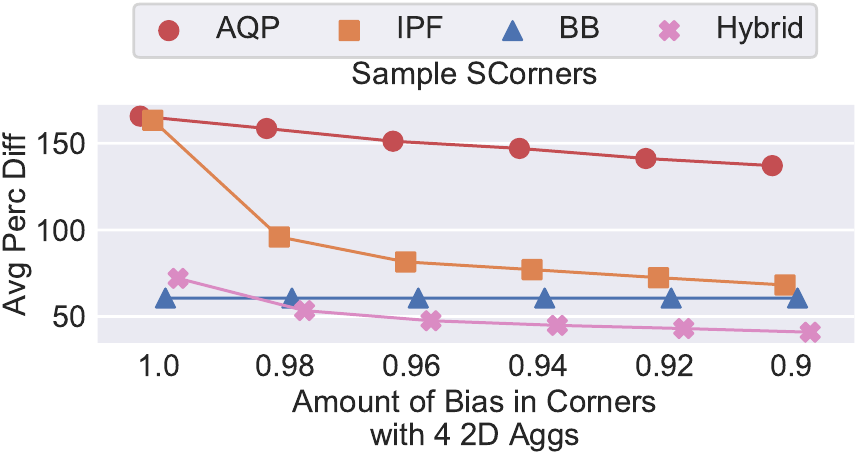}
    \caption{Average percent difference of 100 random point queries for the SCorners using 4 2D aggregates as we decrease the percent bias.}
    \label{fig:flights_increase_bias}
\end{figure}

To examine how the amount of sample bias impacts accuracy, we measure the average percent difference for 100 random point queries using 4 2D aggregates on the \texttt{Flights} sample of Corners as we decrease the percent bias from 100 percent (Corners sample) to 90 percent (SCorners sample), shown in \autoref{fig:flights_increase_bias}.

As soon as the support is the same (bias $< 100$), sample reweighting techniques start performing significantly better. \name's hybrid approach is able to mitigate this difference and performs better than IPF for 100 percent bias.

\begin{figure}[t]
    \centering
    \includegraphics[width=0.9\linewidth]{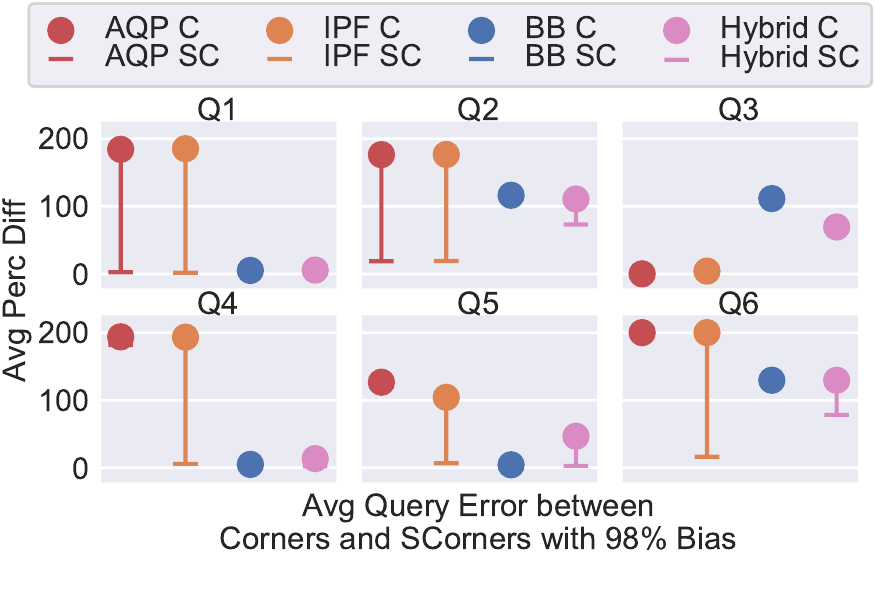}
    \caption{Average query error for six queries between Corners with 100 percent bias (C) and 98 percent bias (SC).}
    \label{fig:increase_bias_queries}
\end{figure}

We further examine the impact of bias on accuracy on more complex SQL queries. Using the benchmark queries in IDEBench~\cite{eichmann2018idebench}, we chose six SQL queries (\autoref{tab:sql_queries}) to show the strengths and weaknesses of \name. We do adapt them to be general \texttt{GROUP BY} queries rather than chained filter queries generated by visualizations, but we maintain the core properties of IDEBench queries that they have an aggregate, zero or more filter predicates, and zero or more joins. With the same setup as \autoref{fig:flights_increase_bias}, we run the queries on Corners with 100 percent bias and 98 percent bias (SCorners) and measure the change in the average percent difference (\autoref{fig:increase_bias_queries}). The queries are run on the post-bucketized data. The circle and horizontal line represent the results on Corners and SCorners, respectively. Note that the horizontal line is sometimes obfuscated by the circle, indicating little to no change in error.
\begin{footnotesize}
\begin{table}
\centering
  \begin{tabular}{|c|p{7.5cm}|}
    \hline
    \textbf{Id} & \textbf{Query} \\ \hline
    1 & \texttt{SELECT O, AVG(E) FROM F} \\ \hline
    2 & \texttt{SELECT O, AVG(E) FROM F WHERE DE = `CA'} \\ \hline
    3 & \texttt{SELECT DE, AVG(E) FROM F WHERE O = `CA'} \\ \hline
    4 & \texttt{SELECT O, COUNT(*) FROM F WHERE E < 120} \\ \hline
    5 & \texttt{SELECT DE, COUNT(*) FROM F WHERE E < 120} \\ \hline
    6 & \texttt{SELECT t.O, s.DE, COUNT(*) FROM F t, F s WHERE f.DE=fs.O AND (f1.DE IN [`CO', `WY'])} \\ \hline
  \end{tabular}
\caption{The six SQL queries run in \autoref{fig:increase_bias_queries}. We leave out the \texttt{GROUP BY} clause and replace \texttt{Flights} with \texttt{F} for space.}
\label{tab:sql_queries}
\end{table}
\end{footnotesize}

These queries highlight a few trends in our techniques. First, all queries except Q3 demonstrate that hybrid and BB outperform the alternatives for 100 percent bias because they miss fewer groups. This does not hold for Q3 because the selection is the same as that for the bias; \ie, CA tuples are already present in the sample. This is also why there is no change in the error between Corners and SCorners.

Second, hybrid and BB perform suboptimal in Q2, Q3, and Q6 because they produce more phantom groups than the number of missed groups in IPF and AQP. As BB is a BN, unless BB has a factor (edge) containing all the attributes in a query, BB cannot know exactly which values exist or not in the population (a known problem with using probability distributions to answer queries~\cite{orr2017probabilistic}). Although BB mitigates this problem by using multiple generated samples to answer queries (\autoref{sec:debias_techniques_model_based_query_answering}), it still produces phantom groups. Q1, Q4, and Q5 are all over attribute pairs that are contained in an edge in BB, meaning BB will not generate phantom groups and have excellent performance.

Lastly, Q6 is a join query looking at flights with a layover in CO or WY. We see that due to phantom tuples, BB and hybrid are not optimal for SCorners, but IPF achieves the lowest error, far surpassing AQP. IPF more accurately rebalances the underrepresented flights leaving CO and WY in the sample.

Finally, we discuss \name in comparison to the technique presented in~\cite{galakatos2017revisiting}. This, as far as we know, is the only AQP technique that does not require access to the population and can be modified to leverage the aggregates (see \autoref{sec:related_work}). However, note that~\cite{galakatos2017revisiting} assumes a normally distributed error in query answers and access to a light hitter index, which does not hold in our setting. As we were unable to get the code, we examine their technique of reformulating the joint probability with conditional probabilities for two attributes as we increase the bias. Their motivation for this query rewrite was to reuse prior query answers, but we can modify it to use aggregate query answers.

With the same setup as for \autoref{fig:flights_increase_bias}, we measure the ratio of hybrid's error over~\cite{galakatos2017revisiting}'s error (\ie $\mathit{err}_{\textrm{\name}}/\mathit{err}_{\textrm{\cite{galakatos2017revisiting}}}$) using one 1D aggregate over \texttt{O}. We issue \texttt{GROUP BY, COUNT(*)} queries over the attribute pairs \texttt{O}-\texttt{DE} and \texttt{DT}-\texttt{DE}. We use hybrid to answer the query directly and, following the technique from~\cite{galakatos2017revisiting}, use the known distribution of \texttt{O} with the conditional probability from the sample to answer the query. The results are shown in~\autoref{tab:perc_improvement_galak}.



\begin{table}
\centering
  \begin{tabular}{|c|c|c|c|c|c|c|}
    \hline
    \textbf{Bias} & 100 & 98 & 96 & 94 & 92 & 90 \\ \hline
    \text{\texttt{O}-\texttt{DE}} &  1.3 & 0.58 & 0.98 & 0.94 & 0.97 & 0.97 \\ \hline
    \text{\texttt{DT}-\texttt{DE}} &  1.0 & 3.8 & 4.8 & 5.3 & 5.6 & 5.7 \\ \hline
  \end{tabular}
\caption{Ratio of \name relative to~\cite{galakatos2017revisiting} for group by queries over \texttt{O}-\texttt{DE} and \texttt{DT}-\texttt{DE}.}
\label{tab:perc_improvement_galak}
\end{table}

For the query over \texttt{O}-\texttt{DE}, \name achieves approximately the same error as \cite{galakatos2017revisiting}. It is, on average, 0.96x the error of \cite{galakatos2017revisiting}. The outlier is for the 98 percent bias where hybrid is 0.58x the error of \cite{galakatos2017revisiting}. This is because hybrid has more phantom groups than the sample has missing groups. For the query over \texttt{DT}-\texttt{DE}, \name achieves the lowest error. \name is able to debias the sample using the aggregate information, while \cite{galakatos2017revisiting} cannot use the information and is equivalent to the standard AQP approach of uniform reweighting. As the number of aggregates increase, \name is able to learn from all of the aggregate information, while \cite{galakatos2017revisiting} must choose which information to use per query.

\subsection{Changing Aggregate Knowledge}
To examine how varying our aggregates impacts accuracy, we run 100 random point queries on two \texttt{Flights} samples and two \texttt{IMDB} samples as we add 1D, 2D, and 3D aggregates. To show the impact of attribute coverage, we add the 1D aggregates in two different orders. For \texttt{Flights}, we add the 1D aggregates in order A---\texttt{F}, \texttt{O}, \texttt{DE}, \texttt{E}, \texttt{DT}---and order B, the reverse of order A. For \texttt{IMDB}, we add the 1D aggregates in order A---\texttt{MY}, \texttt{MC}, \texttt{G}, \texttt{RG}, \texttt{RT}---and order B, the reverse of order A. For both datasets, after adding the 1D aggregates, the 2D and 3D aggregates are added as in \autoref{tab:data_agg}.

\begin{figure}[t]
    \centering
    \includegraphics[width=\linewidth]{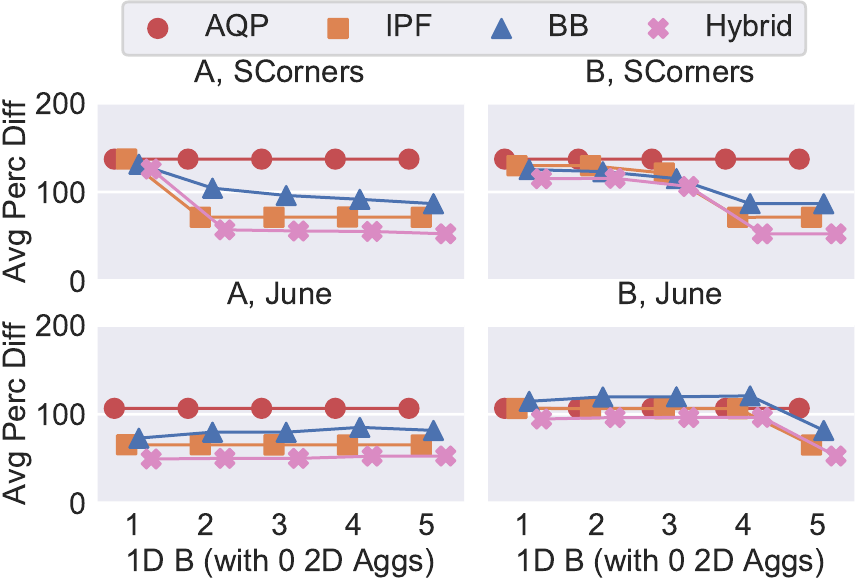}
    \caption{Average percent difference of 100 random point queries for SCorners and June for \texttt{Flights} as more 1D aggregates are added.}
    \label{fig:flights_varying_stat1D}
\end{figure}

\begin{figure}[t]
    \centering
    \includegraphics[width=\linewidth]{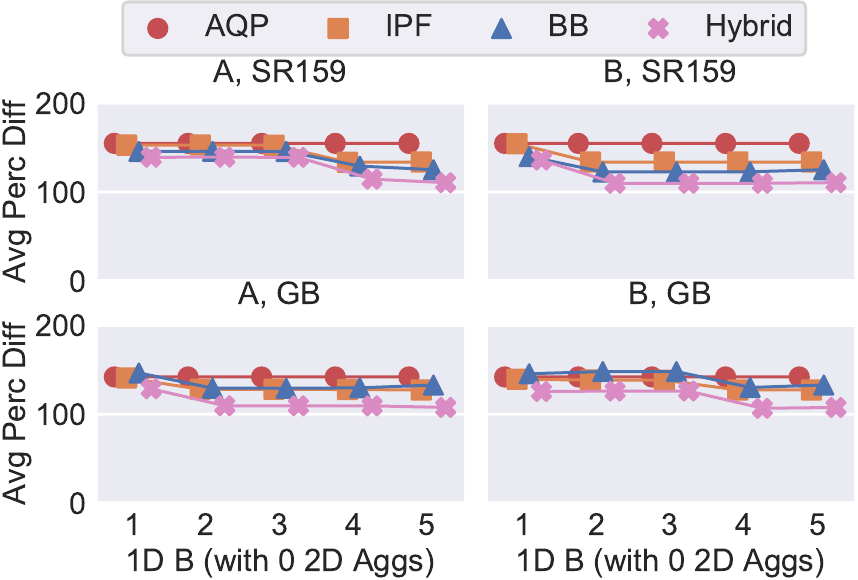}
    \caption{Average percent difference of 100 random point queries for SR159 and GB for \texttt{IMDB} as more 1D aggregates are added.}
    \label{fig:imdb_varying_stat1D}
\end{figure}

\autoref{fig:flights_varying_stat1D} shows a line plot (to better show trends) of the average percent difference for SCorners (top row) and June (bottom row) for order A (left column) and order B (right column). For SCorners, the largest improvement in all \name methods (IPF, BB, and hybrid) is when adding the second attribute in order A (fourth attribute in order B). This is \texttt{O} (the attribute SCorners is biased on) which indicates that \name corrects for the bias the most when the attribute causing the bias is added. As shown in \autoref{fig:imdb_varying_stat1D}, the result is replicated with the June sample and attribute \texttt{F}, with the \texttt{IMDB} sample of SR159 with attribute \texttt{RG}, and with the \texttt{IMDB} sample of GB with attribute \texttt{MC}. Although, the improvement is less pronounced with \texttt{IMDB} as we do not have covering aggregates and the active domain is larger.
\begin{figure}[t]
    \centering
    \includegraphics[width=\linewidth]{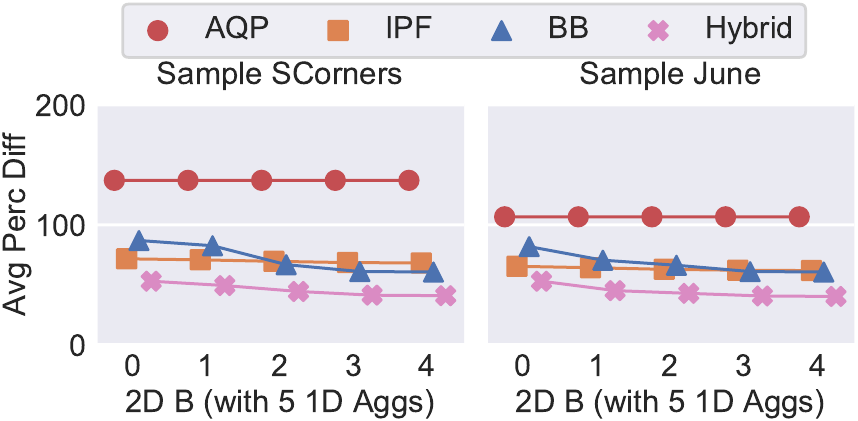}
    \caption{Average percent difference of 100 random point queries for SCorners and June for \texttt{Flights} as more 2D aggregates are added after adding the 5 1D aggregates from \autoref{fig:flights_varying_stat1D}.}
    \label{fig:flights_varying_stat2D}
\end{figure}

\begin{figure}[t]
    \centering
    \includegraphics[width=\linewidth]{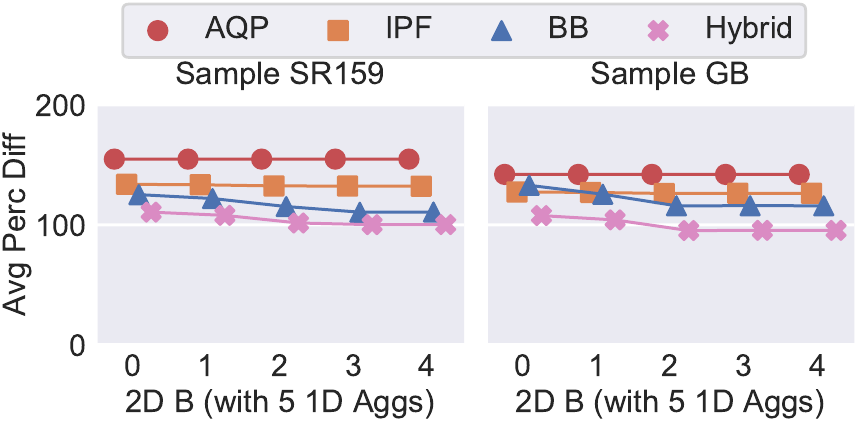}
    \caption{Average percent difference of 100 random point queries for SR159 and GB for \texttt{IMDB} as more 2D aggregates are added after adding the 5 1D aggregates from \autoref{fig:imdb_varying_stat1D}.}
    \label{fig:imdb_varying_stat2D}
\end{figure}

\autoref{fig:flights_varying_stat2D} shows the average percent difference for the same two samples as 2D aggregates are added for \texttt{Flights}. \autoref{fig:imdb_varying_stat2D} shows the same for \texttt{IMDB}. We see that BB improves the most with more aggregates. However, we see diminishing returns after adding 2 aggregates. As more aggregates are added, BB gets closer to hybrid while IPF does not significantly improve with more 2D aggregates.

\begin{figure}[t]
    \centering
    \includegraphics[width=\linewidth]{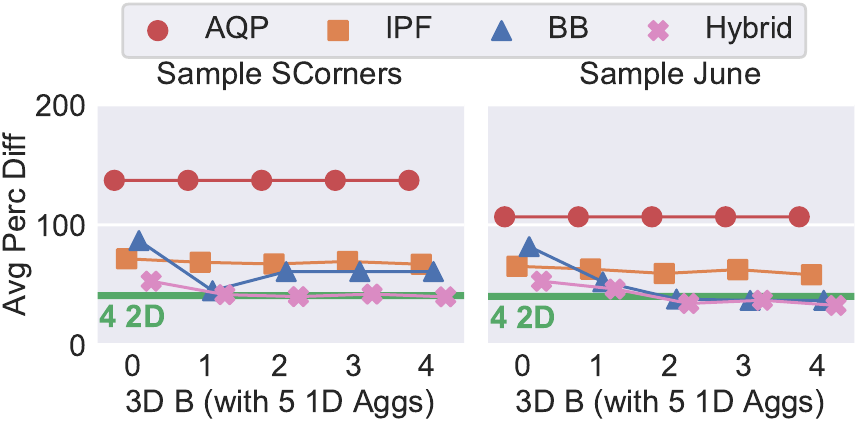}
    \caption{Average percent difference of 100 random point queries for SCorners and Corners for \texttt{Flights} as more 3D aggregates are added after adding the 5 1D aggregates from \autoref{fig:flights_varying_stat1D}.}
    \label{fig:flights_varying_stat3D}
\end{figure}

\begin{figure}[t]
    \centering
    \includegraphics[width=\linewidth]{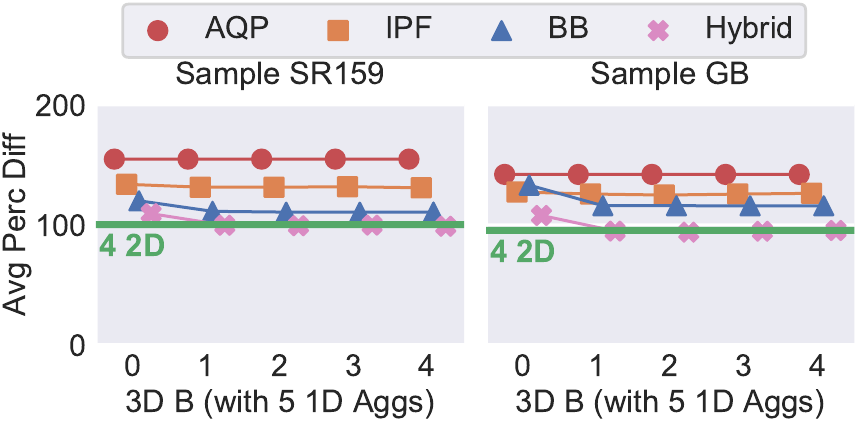}
    \caption{Average percent difference of 100 random point queries for SR159 and R159 for \texttt{IMDB} as more 3D aggregates are added after adding the 5 1D aggregates from \autoref{fig:imdb_varying_stat1D}.}
    \label{fig:imdb_varying_stat3D}
\end{figure}

\autoref{fig:flights_varying_stat3D} shows the average percent difference for the samples SCorners and Corners as 3D aggregates are added after adding 5 1D aggregates for \texttt{Flights}. \autoref{fig:imdb_varying_stat3D} shows the same for \texttt{IMDB} with the samples R159 and SR159. A horizontal green line is added indicating the average percent difference for hybrid after 4 2D aggregates were added.

We see the same trend that aggregates improve BB more than sample reweighting. We further see that adding 3D aggregates causes faster convergence. After adding just a single 3D aggregate are we able to achieve the same error as adding 4 2D aggregates for SCorners.

When looking at \texttt{IMDB}, we notice that adding aggregates does not significantly improve any method. This is due do lacking attribute coverage in the aggregates, especially over dense attributes like \texttt{N} that cause significant error in query answers.

\texttt{Flights}, on the other hand, does have attribute coverage. This means adding 3D aggregates can cause BB and hybrid to have lower error than having 4 2D aggregates, as is shown with June have slightly lower error with 4 3D aggregates than 4 2D aggregates. This trend does not hold with SCorners due to BB learning a less optimal network structure after adding 2 3D aggregates (shown by the dip in the blue line at 1 3D aggregate). With one 3D aggregate only, BB learns that \texttt{O} (origin) should be conditioned on \texttt{DE} (destination) and \texttt{DT} (distance). This makes sense because where a flights starts and how far it goes determines the possible destinations.  With more aggregates, however, the relationship learned is that \texttt{DE} is conditioned on \texttt{DT} and \texttt{O} is conditioned on \texttt{DT}. While similar, this relationship does not accurately represent the underlying distribution as much as the former structure. As our structure learning algorithm is approximate, it will not always learn the optimal structure.

The overall trend is that adding 3D aggregates can improve convergence rates, but does not significantly improve the error for hybrid over just using 2D aggregates.

\subsection{Bayesian Network Performance}
\label{sec:eval_BN}
We explore different BN approaches for structure and parameter learning. When learning a BN, we can use only the sample (S), the aggregates (A), or both (B). Since the aggregates (A) are not always covering, we cannot use it exclusively to learn the BN (unless making suboptimal uniformity assumptions). Therefore, we compare SS (light green), BS (dark green), SB (light blue), and BB (dark blue) where the letters represent whether S or B is used to learn the structure (first) and parameters (second). In addition, we compare against AB (grey) which is when the structure is learned just from $\Gamma$. For attributes not covered by $\Gamma$, we add them as disconnected nodes (uniformity assumption). We measure average percent difference on 100 heavy and light hitter point queries on SCorners while increasing the number of 2D aggregates after adding 5 1D aggregates.

\begin{figure}[t]
    \centering
    \includegraphics[width=0.9\linewidth]{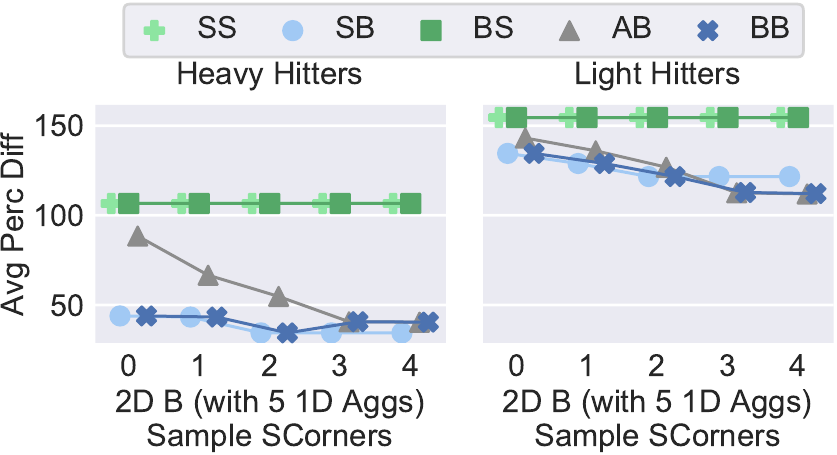}
    \caption{Average percent difference of 100 heavy hitter and light hitter queries over SCorners for the 5 different Bayesian techniques as more 2D aggregates are added with 5 1D aggregates already included.}
    \label{fig:flights_BN_Hitters}
\end{figure}

\autoref{fig:flights_BN_Hitters} shows average percent difference for heavy and light hitters for while increasing the number of 2D aggregates after adding 5 1D aggregates. We see that all approaches perform better for heavy hitters than light hitters and that BB performs best overall. The only time this does not hold is for heavy hitter queries on SCorners once three 2D aggregates have been added. This is due to BB learning a different network structure after adding the aggregate over \texttt{DT} and \texttt{OS}. It adds an edge from \texttt{DT} to \texttt{OS} with three aggregates when it had an edge from \texttt{DE} to \texttt{OS} with only two aggregates. However, the error improvement for light hitter queries for BB after adding three 2D aggregates outweighs the error decrease for heavy hitter queries. As the number of aggregates increases, AB converges to BB because the population information overrides any sample information for structure learning.

Another important trend is that using the sample versus both the aggregates and the sample is more important for parameter learning than for structure learning. SB is more accurate than either BS or SS. Interestingly, BB is more accurate than SB for light hitters but slightly less accurate for heavy hitters. This is due to the fact that greedy structure learning is not guaranteed to be optimal, and SB learns a more accurate network for heavy hitters. We leave it as future work to improve upon structure learning.

\subsection{Sample Reweighting Performance}
We now compare the two different sample reweighting techniques of linear regression (LinReg) and IPF. As we do not see a drastic improvement in adding 2D aggregates when doing sample reweighting, we focus on comparing how they perform on the different \texttt{Flights} samples by measuring error on 100 random point queries using 4 2D aggregates.

\begin{figure}[t]
    \centering
    \includegraphics[width=0.9\linewidth]{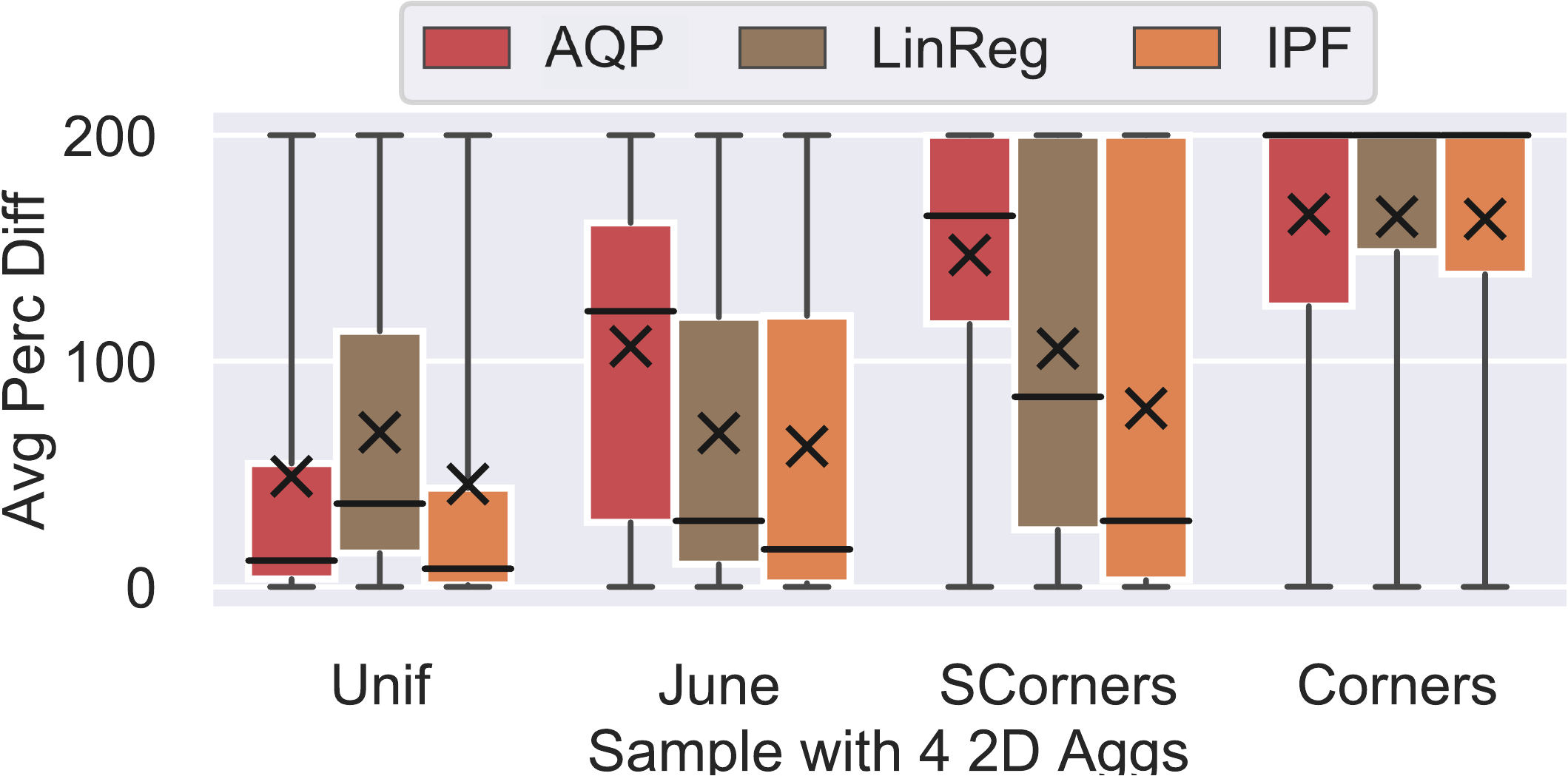}
    \caption{Percent difference of 100 random point queries over four different \texttt{Flights} samples with 4 2D aggregates.}
    \label{fig:linmodels}
\end{figure}

\autoref{fig:linmodels} shows the percent difference of LinReg, IPF, and AQP. Note that AQP does not achieve near zero error on Unif because some of the random point queries are over light hitters which are not in the sample. We see clearly that IPF outperforms LinReg on all cases. While LinReg does outperform AQP in all biased samples, it does not outperform IPF due to correlations in the data. For example, to satisfy aggregates on the \texttt{DT} attribute, LinReg will add weight to the highly correlated attribute values of \texttt{E}. This will help satisfy aggregates on \texttt{DT} but will overall hurt performance because any other tuple with the correlated attribute values will have an incorrect weight.

From \autoref{fig:linmodels}, it may seem that IPF is always superior to LinReg, but when the data is uncorrelated, LinReg should achieve approximately the same error as IPF. Further, if more rows are added to the sample, LinReg does not have to be retrained while IPF does.

\subsection{Pruning Effectiveness}
We use the CHILD dataset to examine our the accuracy of aggregate pruning technique (runtimes as we add aggregates are in \autoref{sec:eval_time}). We use  BB and AB and measure average percent difference for 100 random point queries for 10 randomly chosen attribute sets of size 2, 4, 6, 8, and 10 for a total of 50 attributes sets and 5,000 total point queries. We compare the techniques as we add from 5 to 65 2D aggregates using our pruning technique (Prune) and randomly (Rand) after adding full 1D aggregates. In addition to comparing the error, we also plot the average percent difference if the true Bayesian network is known (opt error).

\begin{figure}[t]
    \centering
    \includegraphics[width=0.9\linewidth]{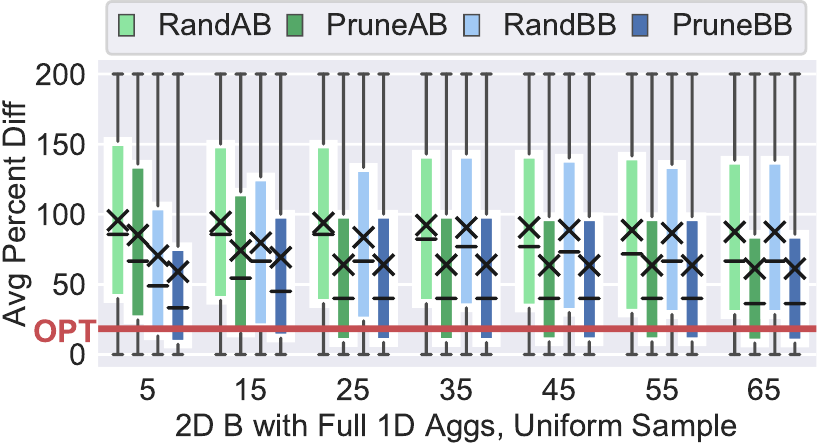}
    \caption{Average percent difference of 100 random point queries using a 10 percent uniform sample of the CHILD dataset with full 1D aggregates as more 2D aggregates are added using Prune and Rand.}
    \label{fig:bn_prune}
\end{figure}

\autoref{fig:bn_prune} shows the average percent difference. The red line shows the median optimal error. We see, again, that BB performs better the AB, especially when fewer aggregates are added. Further, the error with Rand improves more slowly than using Prune to add aggregates. If enough aggregates are added, however, the techniques converge.
\subsection{Execution Time}
\label{sec:eval_time}
Lastly, we examine the query execution time and solver time of the different approaches. We run all timing experiments on the \texttt{IMDB} SR159 sample as \texttt{IMDB} has the larger active domain. For the query execution time, we run 100 random point queries.

\autoref{tab:querytime} shoes the average query execution time of the five different BN methods and any reweighting technique (method of RW). The RW method represents LinReg, IPF, AQP, and a BN generated sample for \texttt{GROUP BY} queries. As the query execution time does not noticeably change as we add more 2D aggregates, we only show results for 4 2D aggregates.
\label{sec:eval_runtime}
\begin{table}[t]
    \centering
    \begin{tabular}{|c|c|c|c|c|c|c|}
    \hline
    Method & RW & SB & AB & SS & BS & BB \\ \hline
    Runtime ($10^{-3}$ s) & 25.3 & 2.49 & 1.97 & 2.45 & 2.26 & 2.07 \\ \hline
    \end{tabular}
    \caption{Average query execution time of 100 random point queries over SR159 with 4 2D aggregates .}
    \label{tab:querytime}
\end{table}

As the main bottle neck to using these methods is the solver time, we report the time it takes to learn the structure for BB (the solver times of the other Bayesian methods are comparable or faster) and the time it takes to learn the parameters of LinReg, IPF, and BB in \autoref{tab:solvertimes} as we increase the number of aggregates.

We see that the structure learning time is negligible compared to the solver time for BB. LinReg is the fastest, followed by IPF, and then by BB. The solver time increases with all methods as we increase the number of 1D aggregates. Surprisingly, as we add more 2D aggregates, the solver time of BB decreases. This is because as we add more 2D constraints to our model, the constraint solver has more direct equality constraints which are instantaneous to solve for.

\begin{table}
\begin{footnotesize}
\centering
    \begin{tabular}{|c|c|c|c|c|c|c|c|c|c|c|}
    \hline
    \multirow{2}{*}{} & \textbf{1D} & 1 & 2 & 3 & 4 & 5 & \multicolumn{4}{c|}{5} \\ \cline{2-11}
                      & \textbf{2D} & \multicolumn{5}{c|}{0} & 1 & 2 & 3 & 4 \\ \hline
    S & BB & 0.3 & 0.3 & 0.4 & 0.7 & 0.7 & 2.3 & 5.3 & 8.4 & 13.2 \\ \hline
    \multirow{3}{*}{P} & Reg & 0.4 & 0.5 & 0.6 & 1.0 & 3.1 & 4.6 & 6.1 & 6.6 & 7.0 \\ \cline{2-11}
                       & IPF & 0.2 & 0.6 & 0.8 & 1.7 & 10.8 & 16.1 & 22.5 & 30.0 & 38.5 \\ \cline{2-11}
                       & BB & 75.1 & 103 & 103 & 161 & 295 & 148 & 68.7 & 63.0 & 58.8 \\ \hline
    \end{tabular}
    \caption{Structure (S) and parameter (P) learning execution times in seconds for LinReg (denoted Reg for space), IPF, and BB using SR159 sample as 1D and 2D aggregates are added.}
    \label{tab:solvertimes}
\end{footnotesize}
\end{table}

\begin{figure}[t]
    \centering
    \includegraphics[width=0.9\linewidth]{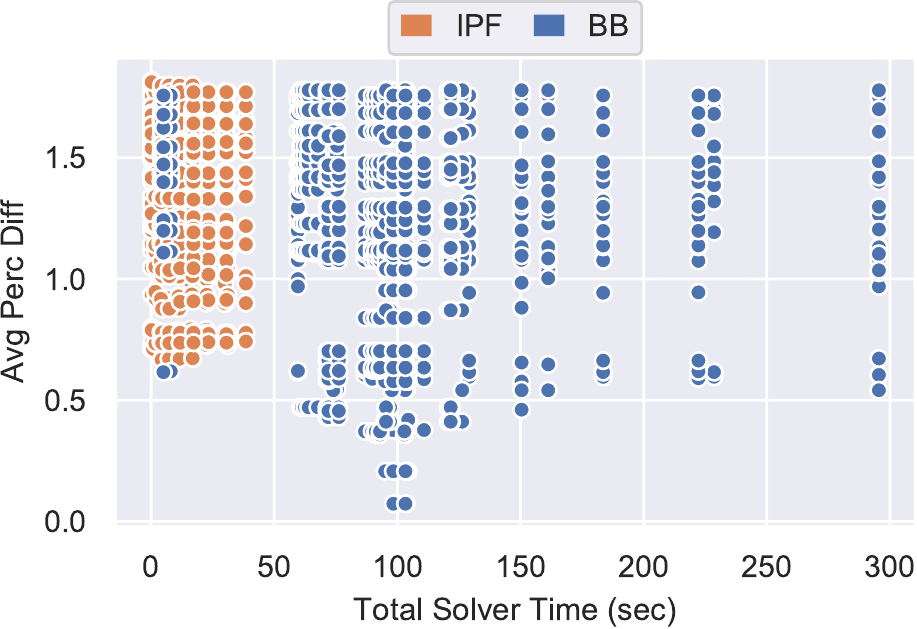}
    \caption{Average percent difference versus total solver time in seconds for IPF and BB on SR159 for various 1D and 2D aggregates.}
    \label{fig:solver_vs_err}
\end{figure}

To show this trend more directly, \autoref{fig:solver_vs_err} shows the average percent difference for 100 random point queries compared to total solver time (structure plus parameter learning) for BB and IPF on SR159 while using different combinations of 1D and 2D aggregates. Specifically, we use our pruning technique to add various combinations of one to four 2D aggregates that cover from one up to five attributes (the 5 attributes of \texttt{MY}, \texttt{MC}, \texttt{G}, \texttt{RG}, \texttt{RT}).
We see that while IPF is almost always faster than BB to solve, BB is capable of achieving lower error. Further, at 100 seconds of solver time, BB achieves the lowest error. This is when the most 2D aggregates are added to the mode which, as previously explained, leads to the fastest solver time.

\section{Related Work}
\label{sec:related_work}
Our technique is related to population synthesis where the goal is to directly generate a population dataset from a sample and either historical population data~\cite{lovell1963seasonal} or aggregate data~\cite{lovelace2015evaluating,beckman1996creating,muller2017generalized,sun2015bayesian,farooq2013simulation,lee2011cross}. \name, however, combines two different techniques, does not assume the sample is representative, and more accurately answers queries over tuples not in the sample.

\name is related to bootstrapping, which is a resampling technique for understanding the uncertainty in queries during AQP~\cite{chaudhuri2017approximate,mozafari2015handbook,li2018approximate,zeng2014analytical}, but in \name, the sample is not representative of the population and the sample probabilities are not known, a requirement for accurate bootstrap.

As discussed in \autoref{sec:introduction}, standard AQP techniques cannot be applied to OWQP. Most related to \name is the work in AQP that reuses past or known results~\cite{peng2018aqp++,galakatos2017revisiting,park2017database}. \cite{peng2018aqp++} uses pre-computed data cube aggregates to improve sample query accuracy, but it is required that the dimensions of the data cube match that of the aggregate queries, which is not true in \name. \cite{galakatos2017revisiting} assumes sample query errors to be normally distribution and requires a light hitter index, and \cite{park2017database} assumes accurate knowledge of the query error. Neither of these conditions hold or are available in \name.

From a data integration standpoint, \name is closely related to answering queries over views (samples)~\cite{lenzerini2002data,halevy2001answering} where the views do not contain complete information. However, we attempt to model the data missing from the data sources whereas data integration deals with knowing when answers are certain or not.

In regards to data cleaning~\cite{ilyas2015trends, chaudhuri2007leveraging, wang2014sample}, while \name is trying to infer missing values from the sample, we are missing entire rows, not just attribute values.

The two related machine learning areas are one class classification~\cite{khan2009survey,goodfellow2014generative} and learning from aggregate labels~\cite{musicant2007supervised, bhowmik2015generalized, chen2006learning}. The main difference is that \name is trying to learn a classifier with aggregate data and does not have aggregate labels of both classes, \ie in the sample and not in the sample.

Our method, in essence, calculates a propensity score for a record~\cite{mccaffrey2013tutorial, rosenbaum1984reducing}. However, standard propensity score techniques require data that is not in the sample to be given, which we do not have. A possible solution is to generate the data outside of the sample~\cite{he2008learning}. We leave this possible technique as future work.

The work of~\cite{chung2018estimating} is a particular subset of our problem. That work tries to take into account unknown unknown values when estimating aggregate queries. Our goal is similar, but we take a machine learning approach to re-weight samples rather than estimating missing values.

\name is similar to~\cite{feldman2015certifying, zemel2013learning} which tries to remove bias from machine learning algorithms, \ie make socially fair classifiers. Our research, however, does not have protected attributes nor does it have access to the entire population.

Considering selection bias and machine learning is the work of~\cite{huang2007correcting,zadrozny2004learning}. \name, however, only has access to the biased test set and does not have sample probabilities to use.


Also using aggregates as constraints is~\cite{vartak2016refinement}. They use aggregates to constrain query answers. The work in~\cite{margaritis2001netcube} discusses using Bayesian networks to approximate a data cube. Our work is similar to both of these except our goal is not to merely answer queries but to also debias a sample.

Similar to how \name treats the biased samples as first class citizens, FactorBase~\cite{qian2015factorbase} and BayesStore~\cite{wang2008bayesstore, wang2011extracting} do the same with graphical models and probabilistic inference. While they both utilize Bayesian networks to model data, their goal is not data debiasing.

The work of~\cite{getoor2001selectivity} uses BNs over relations to do selectivity estimation for queries. \name also uses BNs but is not concerned with multiple relations or selectivity estimation.

\cite{muller2018improved} performs conjunctive query selectivity estimation using both samples and synopses. \name also uses samples and synopses (synopses are population histograms). Our overall goal, however, is to debias the data.

\section{Conclusion}
\label{sec:conclusion}
We introduce a novel query processing paradigm, OWQP. We then present \name, the first prototype OW-DBMS that uses a biased sample and population-level aggregate information to perform OWQP. More importantly, our data debiasing is automatic. \name's hybrid approach merges sample reweighting with population probabilistic modeling to achieve a 70 percent improvement in the median error when compared to uniform reweighting for heavy hitter queries. Further, as shown in \autoref{fig:flights_increase_bias}, \name is robust to differences in the support between the sample and population.

Future work is to extend \name to support continuous data by extending our BN to allow for continuous distributions, as done in~\cite{friedman1998bayesian}. We further want to integrate multiple samples into the debiasing process and investigate alternative techniques to integrate the sample and population into our probabilistic model.

\begin{small}
\noindent\textbf{Acknowledgements.} This work is supported by NSF AITF 1535565, NSF IIS 1907997, and a gift from Intel.
\end{small}

\end{sloppypar}
\small
\bibliographystyle{abbrv}
\bibliography{references}

\end{document}